\documentclass[11pt]{article}
\usepackage{amsmath}
\normalbaselines

\begin{document}
\bibliographystyle{unsrt}

\def\bea*{\begin{eqnarray*}}
\def\eea*{\end{eqnarray*}}
\def\ba{\begin{array}}
\def\ea{\end{array}}
\count1=1
\def\be{\ifnum \count1=0 $$ \else \begin{equation}\fi}
\def\ee{\ifnum\count1=0 $$ \else \end{equation}\fi}
\def\ele(#1){\ifnum\count1=0 \eqno({\bf #1}) $$ \else \label{#1}\end{equation}\fi}
\def\req(#1){\ifnum\count1=0 {\bf #1}\else \ref{#1}\fi}
\def\bea(#1){\ifnum \count1=0   $$ \begin{array}{#1}
\else \begin{equation} \begin{array}{#1} \fi}
\def\eea{\ifnum \count1=0 \end{array} $$
\else  \end{array}\end{equation}\fi}
\def\elea(#1){\ifnum \count1=0 \end{array}\label{#1}\eqno({\bf #1}) $$
\else\end{array}\label{#1}\end{equation}\fi}
\def\cit(#1){
\ifnum\count1=0 {\bf #1} \cite{#1} \else 
\cite{#1}\fi}
\def\bibit(#1){\ifnum\count1=0 \bibitem{#1} [#1    ] \else \bibitem{#1}\fi}
\def\ds{\displaystyle}
\def\hb{\hfill\break}
\def\comment#1{\hb {***** {\em #1} *****}\hb }

\newcommand{\TZ}{\hbox{\bf T}}
\newcommand{\MZ}{\hbox{\bf M}}
\newcommand{\ZZ}{\hbox{\bf Z}}
\newcommand{\NZ}{\hbox{\bf N}}
\newcommand{\RZ}{\hbox{\bf R}}
\newcommand{\CZ}{\,\hbox{\bf C}}
\newcommand{\PZ}{\hbox{\bf P}}
\newcommand{\QZ}{\hbox{\rm eight}}
\newcommand{\HZ}{\hbox{\bf H}}
\newcommand{\EZ}{\hbox{\bf E}}
\newcommand{\GZ}{\,\hbox{\bf G}}

\font\germ=eufm10
\def\goth#1{\hbox{\germ #1}}
\vbox{\vspace{38mm}}

\begin{center}
{\LARGE \bf Bethe Equation of  $\tau^{(2)}$-model and Eigenvalues of Finite-size Transfer Matrix of Chiral Potts Model with Alternating Rapidities} \\[10 mm] 
Shi-shyr Roan \\
{\it Institute of Mathematics \\
Academia Sinica \\  Taipei , Taiwan \\
(email: maroan@gate.sinica.edu.tw ) } \\[25mm]
\end{center}

\begin{abstract}
We establish the Bethe equation of the $\tau^{(2)}$-model in the $N$-state chiral Potts model (including the degenerate selfdual cases) with alternating vertical rapidities.
The eigenvalues of a finite-size transfer matrix of the chiral Potts model are computed by use of functional relations. 
The significance of the "alternating superintegrable" case of the chiral Potts model is discussed, and the degeneracy of $\tau^{(2)}$-model found as in the homogeneous  superintegrable chiral Potts model. 
\end{abstract}
\par \vspace{5mm} \noindent
{\rm 2008 PACS}:  05.50.+q, 03.65.Fd, 75.10.-b \par \noindent

\setcounter{section}{0}

\section{Introduction}
\setcounter{equation}{0}
The theory of $N$-state chiral Potts model is a beautiful and important, but technically difficult, subject in  solvable lattice models.  The lack of difference-property of rapidities inevitably causes a technical complexity in the study of the chiral Potts model, a nature which distinguishes the theory from other known solvable lattice models. Nevertheless, progress has been made on the chiral Potts transfer matrix for the past two decades. For examples, one can study the maximum eigenvalue and low lying excitations of the homogeneous chiral Potts model in the thermodynamic $L \rightarrow \infty$ limit \cite{B91, B90, MR}, calculate the free energy and the interfacial tension \cite{AJP, B94, B03}, and also the eigenvalue spectrum was computed in the superintegrable case  \cite{AMP, B89, B93, B94}. Furthermore, the knowledge was culminated in a recent proof of the order parameter by Baxter \cite{B05}. These results all rely on the technique of functional relations \cite{BBP} about  the transfer matrix and fusion matrices of the associated $\tau^{(2)}$-model  in the extended study of chiral Potts model as a descendent of the six-vertex model found in \cite{BazS}. The study is along the line of $TQ$-relation method invented by Baxter in solving the eigenvalue problem of the eight-vertex model \cite{B72, B73, Bax}. The $Q$-matrix of the  $\tau^{(2)}$-model considered is the chiral Potts transfer matrix. Through the functional-relation approach, the exact results about Onsager-algebra symmetry appeared in the homogeneous superintegrable chiral Potts model \cite{R05o} can serve as a useful model case in the study of symmetry of solvable lattice models, among which are the root-of-unity six-vertex, eight-vertex model \cite{DFM,  FM01, FM02, FM04, F06, R05b, R06Q, R06Q8}, and XXZ chains of higher spin \cite{R06F, R07}. While in the functional-relation symmetry  study of  XXZ chains associated to the cyclic $U_q(sl_2)$ representations labeled by the parameter $\zeta \in \CZ^*$ with $q^N= \zeta^N=1$, these XXZ chains are equivalent to certain chiral Potts models with alternating superintegrable rapidities \cite{R075}. Furthermore, the $Q$-operator investigation of the five-parameter $\tau^{(2)}$-family appearing in \cite{BazS} has shown that the $\tau^{(2)}$-model is indeed the same theory as the chiral Potts model of alternating rapidities when the degenerate forms are included \cite{R0710}. This suggests the results in \cite{AMP, B93, B94, B90, MR} about the eigenvalue spectrum could be better understood through a general setting of the chiral Potts model with alternating rapidities, including the degenerate cases. Indeed,  after sorting out the technical details, one finds the functional relations in \cite{BBP} about the chiral Potts model with alternating rapidities also valid in the degenerate cases (for the details, see \cite{R0710}). Note that in the case of alternating rapidities, Baxter extended the study of chiral Potts  model and functional relations to a general $\tau^{(2)}$-model in \cite{B04}, where the vertical rapidities are not necessarily required in the same curve. The Baxter's "inhomogeneous" model is more general than the $\tau^{(2)}$-model in this work where we assume all rapidities in the same rapidity curve.  For convenience, throughout this paper by the $N$-state chiral Potts model (CPM),  we always mean the chiral Potts model with alternating rapidities and degenerate cases included,  unless otherwise stated. The purpose of this paper is to compute the eigenvalue spectrum of the finite(-size) transfer matrix of CPM. Note that the solution of the homogenous CPM, except in the superintegrable case, is yet unknown to the best of the author's knowledge. It is a challenge to obtain all the eigenvalues as they can be solved completely for the  finite-size Ising model \cite{Bax, O}. In the present paper, we find an explicit formula of eigenvalues of the finite transfer matrix of CPM by use of functional relations in \cite{BBP} and the Bethe equation of $\tau^{(2)}$-matrix generalized as in \cite{MR}.  We first obtain the $\tau^{(2)}$-eigenvalues using  the solutions of Bethe equation established by the Wiener-Hopf splitting method in \cite{MR}.  Those Bethe solutions are formed compatibly with the $\tau^{(2)}T$-relation in CPM. We then find that 
the scheme in \cite{B90, MR} for the investigation of  maximum eigenvalue and excitations  of large lattice limit in the homogeneous CPM can be extended to our study of solving all eigenvalues of  the finite chiral Potts transfer matrix. Here, aside the conceptual precision, technical accuracy is also needed for the correct expression of these eigenvalues and quantum numbers as the exact form is required to reproduce results in homogeneous superintegrable CPM previously known in \cite{AMP, B93, B94}. This paper is organized as follows. In section \ref{sec:TauCP}, we recall some known results in $\tau^{(2)}$-model and CPM. In section \ref{ssec.tau2}, we first summarize some basic facts about the fusion relation of $\tau^{(2)}$-model (for more details, see e.g. \cite{R06F, R0710}). Then we formulate various forms of rapidity parameters in CPM and introduce necessary notations for later use. In section \ref{ssec.CP}, we recall briefly the main definitions of chiral Potts transfer matrix and its relation with $\tau^{(2)}$-matrix in \cite{AMP, BBP}.  
In section \ref{sec:Bethe}, we discuss the Bethe equation of $\tau^{(2)}$-matrix in CPM. Suggested by the $\tau^{(2)}T$-relation in CPM, we describe in section \ref{ssec.t2TB} a general mechanism of constructing $\tau^{(2)}$-eigenvalues as solutions of the fusion relation. Then in section \ref{ssec.Bethe}, we apply the Wiener-Hopf splitting technique in \cite{MR} to establish the Bethe equation of $\tau^{(2)}$-model. 
In section \ref{sec:EiCH}, we compute the eigenvalues of the finite transfer matrix in CPM. Here we follow \cite{B90} and define the normalized chiral Potts transfer matrix with alternating rapidity parameters. First in section \ref{ssec.FnRel}, we recall and address technical details about functional relations in terms of the normalized chiral Potts transfer matrix. Those formulas could be known for specialists as a straitforward extension of the homogeneous case  \cite{B90}. However, the explicit formulas of CPM (including the degenerate case) in the alternating-rapidity case have not been previously published in the literature to the best of our knowledge. Since the correct expressions will be needed in later discussions, we present a derivation of those relations here in spite of much of the result just paraphrasing the work of \cite{B90} in a slightly general form. In section \ref{ssec.EigV}, we derive the formula of eigenvalues of the finite chiral Potts normalized transfer matrix by use of functional relations and results in section \ref{ssec.Bethe} about Bethe solutions of  $\tau^{(2)}$-model. In section \ref{ssec.AlSup}, we discuss the "alternating superintegrable" case, where a simplification in the Bethe equation occurs, so is the expression of eigenvalues of $\tau^{(j)}$ and chiral Potts transfer matrix. The degeneracy of the $\tau^{(2)}$-matrix is found, and a comparison of the result with that in the homogeneous CPM is also given there. Finally we close in section \ref{sec.F} with some concluding remarks.

\section{The  $\tau^{(2)}$-model and the $N$-state Chiral Potts Model \label{sec:TauCP}}
\setcounter{equation}{0}
In this paper, $\CZ^N$ denotes the vector space of $N$-cyclic vectors with the basis $|n \rangle, n \in \ZZ_N ~ (:= \ZZ/N\ZZ)$. We fix the $N$th root of unity $\omega = e^{\frac{2 \pi {\rm i}}{N}}$, and  $X, Z$, the Weyl $\CZ^N$-operators :
$$
 X |n \rangle = | n +1 \rangle , ~ \ ~ Z |n \rangle = \omega^n |n \rangle ~ ~ \ ~ ~ (n \in \ZZ_N) ,
$$
satisfying  the relations $XZ= \omega^{-1}ZX$ and  $X^N=Z^N=1$.
\subsection{The $\tau^{(2)}$-model \label{ssec.tau2}}
The $\tau^{(2)}$-model (also called the Baxter-Bazhanov-Stoganov model \cite{GIPS, GIPST1, GIPST2}) is the five-parameter family associated to the $L$-operators of $\CZ^2$-auxiliary, $\CZ^N$-quantum space with entries expressed by the Weyl operators $X, Z$:
\be
{\tt L} ( t ) = \left( \begin{array}{cc}
        1  -  t \frac{{\sf c}  }{\sf b' b} X   & (\frac{1}{\sf b }  -\omega   \frac{\sf a c }{\sf b' b} X) Z \\
       - t ( \frac{1}{\sf b'}  -   \frac{\sf a' c}{\sf b' b} X )Z^{-1} & - t \frac{1}{\sf b' b} + \omega   \frac{\sf a' a c }{\sf b' b} X
\end{array} \right) ,
\ele(L)
satisfying the Yang Baxter equation for the asymmetry six-vertex $R$-matrix. Here $t$ is the spectral variable, and ${\sf a, b, a', b', c}$ are non-zero complex parameters. The monodromy matrix of the chain size $L$ is  
\be
\bigotimes_{\ell=1}^L  {\tt L}_\ell (t)  =  \left( \begin{array}{cc} A_L(t)  & B_L (t) \\
      C_L (t) & D_L(t)
\end{array} \right), \ \ {\tt L}_\ell (t)= {\tt L}(t) \ {\rm at \ site} \ \ell,
\ele(monM)
The $\tau^{(2)}$-matrix is the $\omega$-twisted trace of the above monodromy matrix:
\be
\tau^{(2)}(t) = A_L(\omega t) + D_L(\omega t),  
\ele(tau2) 
which form a commuting family of $\stackrel{L}{\otimes} \CZ^N$-operators, commuting with the spin-shift operator, denoted again by $X \ (:= \prod_{\ell} X_\ell) $ when no confusion could arise.  The $\tau^{(2)}$-eigenfunction is a polynomial in $t$ of degree $L$, and its $X$-eigenvalue defines the charge $\omega^Q$ with $Q \in \ZZ_N$. It is known that 
the quantum determinant of (\req(monM)) is equal to $z(t) X$ where 
\be z(t) = ( \frac{\omega {\sf c}({\sf a b}- t)({\sf a' b'}- t)}{({\sf b' b})^2})^L , 
\ele(zt)
and the "classical" monodromy matrix is 
\bea(l) 
\langle \bigotimes_{\ell=1}^L  {\tt L}_\ell \rangle = \left( \begin{array}{cc} \langle A_L \rangle  & \langle B_L \rangle \\
      \langle C_L \rangle & \langle D_L\rangle
\end{array} \right)= \langle {\tt L}_1 \rangle \langle {\tt L}_2 \rangle \cdots \langle {\tt L}_L \rangle  (= \langle {\tt L} \rangle^L) , \\
  \langle {\tt L} \rangle  = \frac{1}{{\sf b'}^N {\sf b}^N } \left( \begin{array}{cc}
        {\sf b'}^N {\sf b}^N  -  {\sf c}^N t^N  & {\sf b'}^N  - {\sf a}^N {\sf c}^N  \\
        - ( {\sf b}^N - {\sf a'}^N {\sf c}^N )t^N & {\sf a'}^N {\sf a}^N {\sf c}^N - t^N 
\end{array} \right)   
\elea(avM)
(\cite{GIPS} (88) (45), \cite{R06F} (2.9) (2.24),  \cite{Tar}).
Here $\langle O \rangle \ := \prod_{i=0}^{N-1} O(\omega^i t)$ denotes the average of the (commuting family of) operators $O(t)$ for $t \in \CZ$. From the $L$-operator (\req(L)), one can construct $\tau^{(j)}$-matrices, with $\tau^{(0)}=0, \tau^{(1)}= I$ and $\tau^{(2)}$ in (\req(tau2)), so that the fusion relation holds:
\bea(ll)
{\rm Recursive \ ~ relation}:&\tau^{(2)}(\omega^{j-1} t) \tau^{(j)}(t) =  z( \omega^{j-1} t) X \tau^{(j-1)}(t)  + \tau^{(j+1)}(t) , \ \ j \geq 1 ; \\
{\rm Boundary \ ~ relation}:&\tau^{(N+1)}(t) = z(t) X \tau^{(N-1)}(\omega t) + u(t) I ,
\elea(fus)
where $z(t)$  is in (\req(zt)), and $u(t)= \langle A_L \rangle + \langle D_L \rangle (= $ the trace of $\langle {\tt L} \rangle^L)$  (see, \cite{R06F} Proposition 2.1, \cite{GIPS} (107), \cite{R06F} (2.30) and  \cite{R0710} (2.25)). By the recursion relation in 
(\req(fus)), $\tau^{(j)} (t)$ can be expressed as a $\tau^{(2)}$-"polynomial"  of degree $(j-1)$ for $ j \leq N+1 $,  
\begin{eqnarray*}
\tau^{(j)} (t)  = \prod_{s=0}^{j-2} \tau^{(2)}(\omega^s t) + \sum_{k=1}^{[\frac{j-1}{2}] } (-X)^k  \sum_{1 \leq i_1 <'i_2 <' \cdots <' i_k \leq j-2}\prod_{\ell=1}^k \bigg( \frac{z(\omega^{i_\ell} t)}{\tau^{(2)}(\omega^{i_\ell-1 }t )\tau^{(2)}(\omega^{i_\ell }t)} \prod_{s=0}^{j-2} \tau^{(2)}(\omega^s t) \bigg) 
\end{eqnarray*}
where the notion $i_\ell <' i_{\ell+1}$ means $i_\ell + 1 < i_{\ell+1}$. The boundary relation in (\req(fus)) then gives rise to the functional equation of $\tau^{(2)}(t)$: 
\begin{eqnarray}
 \prod_{s =0}^{N-1} \tau^{(2)}(\omega^s t) + \sum_{k=1}^{[\frac{N}{2}] } (-X)^k  \sum_{ I_k  \in {\cal I}_k } \prod_{i \in I_k} \bigg( \frac{z(\omega^{i} t)}{\tau^{(2)}(\omega^{i-1 }t)\tau^{(2)}(\omega^{i }t)} \prod_{s=0}^{N-1} \tau^{(2)}(\omega^s t) \bigg) = u ( t ) I \ , \label{tau2eq}
\end{eqnarray}
where the index sets ${\cal I}_k$ run through all subsets $I_k$ of $\ZZ_N$ with $k$ distinct elements such that $i \not\equiv i'+1 \pmod{N}$ for $i , i' \in I_k$. For examples, the expressions of (\ref{tau2eq}) for $N= 3, 4$ are
$$
\begin{array}{ll}
N=3: & \prod_{j=0}^2\tau^{(2)}(\omega^j t) -  ( \sum_{j=0}^2 z(\omega^j t)\tau^{(2)}(\omega^{j+1} t) )  X = u ( t ) I ; \\
N=4: & \prod_{j=0}^3 \tau^{(2)}(\omega^j t) - ( \sum_{j=0}^3 z(\omega^j t)\tau^{(2)}(\omega^{j+1} t) )X  + (z(t) z(\omega^2 t) + z(\omega t)z(\omega^3 t) )X^2 = u ( t ) I ,
\end{array}
$$
(see \cite{R04} (13)-(15)).

There  is a three-parameter subfamily of $\tau^{(2)}$-models, denoted by $\tau^{(2)}_{p, p'}$,  appeared in the $N$-state chiral Potts model with alternating vertical rapidities $p, p'$, whose coordinates $(x, y, \mu) \in \CZ^3$ satisfy either an algebraic curve relation for the parameter $k'$: 
\be
{\goth W}_{k'}: ~ ~ k x^N  = 1 -  k'\mu^{-N}, \ ~ \  k  y^N  = 1 -  k'\mu^N , \ ~ \ ( k' \neq \pm 1, 0, \ k^2 + k'^2 = 1 ), 
\ele(xymu)
(see, e.g. \cite{BBP}) or the various degenerate $k' = \pm 1$ versions of the above curves, defined by one of the following equations (\cite{R0710} (3.22) (3.25) (4.3)):
\bea(ll)
{\goth W}_1^\prime :&x^N = 1- \mu^{-N}, \  y^N = 1- \mu^N ; \\
{\goth W}_1^{\prime \prime} :& x^N + y^N = 1, \ \mu^N=1 ; \\
{\goth W}_1^{\prime \prime \prime} :& x^N + y^N = 0, \ \mu^N = \pm 1  . \\
\elea(Cdeg)
The genus of a rapidity curve in (\req(xymu)) (\req(Cdeg)) is $(N^3-2N^2+1), \frac{N^3-3N^2 +2 }{2}, \frac{N^2-3N +2 }{2}, 0$ respectively, and each is invariant under the automorphisms:
\bea(ll)
R : (x, y, \mu ) \mapsto (y, \omega x, \mu^{-1} ) , &
T : (x, y, \mu ) \mapsto (\omega x, \omega^{-1} y, \omega^{-1}\mu ) ,\\
U : (x, y, \mu ) \mapsto (\omega x, y, \mu ), & U' ~(= R^2U^{-1}): (x, y, \mu ) \mapsto (x, \omega y, \mu ). 
\elea(Aut)
We shall use $t, \lambda, {\bf x}$ to denote the variables
\be 
t= xy, ~ ~ ~ \lambda = \mu^N , \ \ {\bf x} = x^N . 
\ele(tlda)
By eliminating $\lambda$, a curve in (\req(xymu)) or (\req(Cdeg)) is reduced to a $xy$-curve correspondingly:
\bea(llll)
x^N + y^N = k(1 + x^N y^N), & x^N + y^N =  x^N y^N , & 
x^N + y^N = 1, & x^N + y^N = 0 .
\elea(xyc)
The quotient of the above $xy$-curve by the automorphism $T$ in (\req(Aut)) depends on the variables $t$ and ${\bf x}$, which defines a hyperelliptic curve $W$ of genus $N-1, [\frac{N-1}{2}], [\frac{N-1}{2}]$ and $0 $ respectively:
\bea(lll)
W = &W_{k'} :  t^N = \frac{(1- k' \lambda )( 1 - k' \lambda^{-1}) }{k^2 } , & 
W_1^\prime :  t^N= (1- \lambda)(1-\lambda^{-1} ) ,  \\
& W_1^{\prime \prime} : t^N = {\bf x} (1- {\bf x}) , &
W_1^{\prime \prime \prime} : t^N =  - {\bf x}^2 .
\elea(hW)
Hereafter in the case $W_1^{\prime \prime \prime}$, we shall consider only the odd $N$  as for even $N$ it consists of two rational irreducible components. We will use $\eta$ to denote the complex number 
\be
 \eta := (\frac{1-k'}{1+k'})^\frac{1}{N} , ~  4^\frac{1}{N} , ~ 4^\frac{-1}{N}, ~ 0 
\ele(eta)
respectively for $W_{k'}, W_1^\prime, W_1^{\prime \prime}, W_1^{\prime \prime \prime}$ in (\req(hW)).
The  branched locus in the cases of $W_{k'}$ and  $W_1^{\prime \prime \prime}$ is defined by  $t^N = \eta^{\pm N}$. For $W_1^\prime$ and $W_1^{\prime \prime}$, the branched value is: $t^N =\eta^N$,  while for odd $N$, there is another branched value with $t = 0, \infty$ respectively. 
We denote $t_0^+, t_0^-$ the following branched values:
\be
(t_0^+, t_0^-) := (\eta^{-1}, \eta) ~ ~ {\rm for} ~ W_{k'} \ {\rm and} \ W_1^{\prime \prime \prime}, ~ ~ ~ ~ (\eta ,\eta \omega) ~ ~ {\rm for} ~ W_1^\prime \ {\rm and} \  W_1^{\prime \prime}. 
\ele(t0)
For later convenience, we shall use $(t, \sigma)$ to denote coordinates of $W$ in (\req(hW)) together with the conjugate variable $\sigma^\dagger$ of $\sigma$:
\bea(lll)
\sigma = \lambda, & \sigma^\dagger= \lambda^{-1},  & (t, \lambda) \in W_{k'} \ {\rm or} \ ~ W_1^\prime ; \\
\sigma = {\bf x} , & \sigma^\dagger= 1- {\bf x}, \ - {\bf x}, & (t, {\bf x}) \in W_1^{\prime \prime} \ {\rm or} \ ~  W_1^{\prime \prime \prime} \ {\rm respectively}.
\elea(sigma)
The interchange of $x^N$ and $y^N$ of a  $xy$-curve in (\req(xyc)) induces the hyperelliptic involution of $W$ in (\req(hW)):  $(t, \sigma ) \mapsto (t, \sigma^\dagger )$. Hereafter, the letters $p, q, \ldots$ will denote rapidities in a curve in (\req(xymu)) or (\req(Cdeg)), and write  their coordinates by $x_p, y_p, \mu_p, t_p, \lambda_p $ whenever it will be necessary to specify the element $p$. The $L$-operator of $\tau^{(2)}_{p, p'}(t)$ in (\req(L)) is defined by the affine coordinates of $p, p'$ in (\req(xymu)) or (\req(Cdeg))  with the parameters
\be
({\sf a, b, a', b', c}) = (x_p, y_p, x_{p'}, y_{p'}, \mu_p \mu_{p'}) ,
\ele(tpp')
(see \cite{BBP} (3.44a), \cite{R0710} (2.18) (3.23) (4.4)), and the spectral parameter $t$ is related to the rapidity variable $q$ in (\req(xymu)) or (\req(Cdeg)) by (\req(tlda)). 
The $z(t)$ in (\req(fus)) becomes 
\be
z (t)= ( \frac{\omega \mu_p \mu_{p'}(t_p-t)(t_{p'}-t)}{y_p^2 y_{p'}^2} )^L ,
\ele(zCP)
and $u(t)$ in (\req(fus)) is  now expressed by $u(t)= \alpha_q + \overline{\alpha}_q$, where $\alpha_q$, $\overline{\alpha}_q$ are the eigenvalues of $\langle {\tt L} \rangle^L$ in (\req(avM)) (\cite{BBP} (4.28) (4.29), \cite{R0710} (2.26) and section 5):
\bea(lll)
\alpha_q & = ( \frac{(t_p^N-t^N)(y_{p'}^N-x^N)}{y_p^N y_{p'}^N(x_p^N-x^N)})^L & = ( \frac{(t_{p'}^N-t^N)(y_p^N-x^N)}{y_p^N y_{p'}^N(x_{p'}^N-x^N)})^L = \alpha (\sigma) , \\ 
 \overline{\alpha}_q & = ( \frac{(t_p^N-t^N)(y_{p'}^N-y^N)}{y_p^N y_{p'}^N(x_p^N-y^N)})^L & = ( \frac{(t_{p'}^N-t^N) (y_p^N-y^N)}{y_p^N y_{p'}^N(x_{p'}^N-y^N)})^L  = \alpha (\sigma^\dagger),
\elea(aaq)
with the variables $\sigma, \sigma^\dagger$ in (\req(sigma)). 
Note that $\alpha_q, \overline{\alpha}_q$ are related to $z(t)$ by the equality
\be
\alpha_q \overline{\alpha}_q = z(t)z(\omega t) \cdots z(\omega^{N-1}t),
\ele(aaz)
which is the connection between the determinant of $\langle \bigotimes_{\ell=1}^L {\tt L}_\ell \rangle $ and quantum determinant of $\bigotimes_{\ell=1}^L {\tt L}_\ell$:  
${\rm det}  \langle \bigotimes_{\ell=1}^L {\tt L}_\ell \rangle =  \langle {\rm det}_q \bigotimes_{\ell=1}^L {\tt L}_\ell  \rangle$. It is shown in \cite{R0710} that one can always reduce a $\tau^{(2)}$-model to a chiral Potts $\tau^{(2)}_{p, p'}$ through a procedure of a special gauge transform and the rescaling of spectral parameters (\cite{R0710} (2.21) and (2.22)):
$$
({\sf a, b, a', b', c}) \mapsto (\lambda \nu^{-1} {\sf a}, \nu {\sf b}, \nu {\sf a}', \lambda \nu^{-1} {\sf b}', {\sf c} ), \ \ \nu, \lambda \in \CZ^*.
$$
Indeed,  a $\tau^{(2)}$-model whose the parameters satisfy 
$$
({\sf c}^N {\sf a'}^N- {\sf b}^N)( {\sf c}^N {\sf a}^N -{\sf b'}^N)({\sf a'}^N- {\sf b}^N)( {\sf a}^N -{\sf b'}^N) \neq 0 , 
$$
is equivalent to a $\tau^{(2)}_{p, p'}$. The rest $\tau^{(2)}$-models  are either with the pseudovacuum state where the algebraic Bethe technique applies, or the "zero-temperature" ($k'=0$) limit of CPM (see, \cite{R0710} (3.20) and section 4.2). For the rest of this paper, we shall only consider the $\tau^{(2)}$-model with the above constraint, and write 
$$
\tau^{(2)}(t)  = \tau^{(2)}_{p, p'} (t) , \ \ t \in \CZ ,
$$
with rapidities $p, p'$  in a curve in (\req(xymu)) or (\req(Cdeg)).

\subsection{ The chiral Potts model and the degenerate selfdual model \label{ssec.CP}}
Using the coordinates $(x , y , \mu )$ of rapidities $p, q$ in (\req(xymu)) or (\req(Cdeg)), one defines the Boltzmann weights of the CPM by
\be
\frac{W_{p,q}(n)}{W_{p,q}(0)}  = (\frac{\mu_p}{\mu_q})^n \prod_{j=1}^n
\frac{y_q-\omega^j x_p}{y_p- \omega^j x_q }  , \ ~ \
\frac{\overline{W}_{p,q}(n)}{\overline{W}_{p,q}(0)}  = ( \mu_p\mu_q)^n \prod_{j=1}^n \frac{\omega x_p - \omega^j x_q }{ y_q- \omega^j y_p }.  
\ele(Weig)
The rapidity condition ensures the $N$-periodic property of the above Boltzmann weights for integers $n$. For convenience, we shall assume $W_{p,q}(0)= \overline{W}_{p,q}(0) =1$ without loss of generality.  Let $\overline{W}^f_{pq}(n)$ be the Fourier transform of $\overline{W}_{pq}(n)$  (\cite{BBP} (2.22)-(2.24)):
\bea(ll)
\overline{W}^f_{pq}(n)= \sum_{k=0}^{N-1} \omega^{nk} \overline{W}_{pq}(k), & \frac{\overline{W}^f_{pq}(n)}{\overline{W}^f_{pq}(0)} = \prod_{j=1}^n \frac{y_q - \omega^j \mu_p\mu_q  x_p}{y_p  - 
\omega^j \mu_p\mu_q x_q} .
\elea(Wf)
Denote 
\bea(lll)
g_p(q)&: =  \prod_{n=0}^{N-1} W_{pq}(n)& = (\frac{\mu_p}{\mu_q})^{\frac{(N-1)N}{2}} \prod_{j=1}^{N-1} (\frac{  x_p -\omega^j y_q }{  x_q- \omega^j y_p })^j, \\
\overline{g}_p (q)&: = {\rm det}_N(\overline{W}_{p q}(i-j))&(= \prod_{n=0}^{N-1} \overline{W}^f_{pq}(n)) = (\overline{W}^f_{pq}(0))^N \prod_{j=1}^{N-1} (\frac{ \mu_p\mu_q  x_p - \omega^j y_q }{\mu_p\mu_q x_q - \omega^j y_p  })^j .
\elea(gg1)
Using (\req(Wf)), one finds the identity  (\cite{BBP} (2.44))\footnote{The identity (\req(g-q)) is formula (2.44) in \cite{BBP} where rapidities are in (\req(xymu)). The same equality holds also in the degenerate model with rapidities in (\req(Cdeg)). }:
 \be
\overline{g}_p (q) = N^{\frac{N}{2}} {\rm e}^{{\rm i} \pi (N-1)(N-2)/12} \prod_{j=1}^{N-1} \frac{(t_p - \omega^j t_q)^j }{(x_p - \omega^j x_q)^j (y_p - \omega^j y_q)^j }. 
\ele(g-q)
It is known  \cite{AMPT, AuP, BPA, FatZ,  MaS, MPTS} that the Boltzmann weights in (\req(Weig)) satisfy the star-triangle relation 
\be
\sum_{n=0}^{N-1} \overline{W}_{qr}(j' - n) W_{pr}(j - n) \overline{W}_{pq}(n - j'')= R_{pqr} W_{pq}(j - j')\overline{W}_{pr}(j' - j'') W_{qr}(j -j'')    
\ele(TArel)
where $R_{pqr}= \frac{f_{pq}f_{qr}}{f_{pr}}$ with $f_{pq} =  \frac{\overline{g}_p (q)^{1/N} }{g_p(q)^{1/N}}$. 
On a lattice of the horizontal size $L$, the combined weights of
intersections with vertical rapidities $p, p'$ between two consecutive rows define the $\stackrel{L}{\otimes} \CZ^N$-operator, called the chiral Potts transfer matrix, commuting with the spin-shift operator $X$:
\bea(l)
T_{p,p'} (q)_{\{j \}, \{j'\}} = \prod_{\ell =1}^L W_{p,q}(j_\ell - j'_\ell )
\overline{W}_{p',q}(j_{\ell+1} - j'_\ell), \\
\widehat{T}_{p, p'} (q)_{\{j \}, \{j'\}} = \prod_{\ell =1}^L \overline{W}_{p,q}(j_\ell - j'_\ell) W_{p',q}(j_\ell - j'_{\ell+1}) .
\elea(Tpq)
Here $q$ is an arbitrary rapidity, and  $j_\ell, j'_\ell \in \ZZ_N$ with the periodic condition imposed by defining $L+1=1$. 
As in \cite{BBP} (2.39)-(2.42),  the operators in (\req(Tpq)) are indeed single-valued in $x_q$ and $y_q$: $T_{p, p'}(q)= T_{p, p'}(x_q, y_q)$, $\widehat{T}_{p, p'}(q)= \widehat{T}_{p, p'}(x_q, y_q)$, satisfying the relations 
\bea(l)
T_{p, p'}(\omega x_q, \omega^{-1} y_q)= (\frac{ ( y_p- \omega x_q ) (y_{p'} - \omega^{-1} y_q)}{\mu_p  \mu_{p'} (\omega x_p -  y_q) (x_{p'} -  x_q)  })^L X^{-1} T_{p,p'}(x_q, y_q) , \\ 
\widehat{T}_{p, p'}(\omega x_q, \omega^{-1} y_q)= (\frac{ ( y_{p'}- \omega x_q ) (y_p - \omega^{-1} y_q)}{\mu_p  \mu_{p'} (\omega x_{p'} -  y_q) (x_p -  x_q)  })^L X^{-1} \widehat{T}_{p,p'}(x_q, y_q) . 
\elea(Tqrel)
The star-triangle relation (\req(TArel)) yields the commutative relation for rapidities $q$ and $r$:
\be
T_{p, p'} (q) \widehat{T}_{p, p'} (r) = (\frac{f_{p'q}f_{pr}}{f_{pq}f_{p'r}})^L T_{p, p'} (r) \widehat{T}_{p, p'} (q) , \ \ \widehat{T}_{p, p'} (q) T_{p, p'} (r) = (\frac{f_{pq}f_{p'r}}{f_{p'q}f_{pr}})^L  \widehat{T}_{p, p'} (r) T_{p, p'} (q) .
\ele(TTc)
The commuting family of $Q$-operators is defined by 
$$
Q_{p, p'} (q) = \widehat{T}_{p, p'} (q_0)^{-1}  \widehat{T}_{p, p'} (q) = (\frac{f_{pq}f_{p'q_0}}{f_{p'q}f_{p q_0}})^L  T_{p, p'} (q) T_{p, p'} (q_0)^{-1}  \ (q \in {\goth W}_{k'})
$$
where $q_0$ is an element in ${\goth W}_{k'}$ with non-singular $\widehat{T}_{p, p'} (q_0)$ and $T_{p, p'} (q_0)$. Note that for  $p \neq p'$, the commutativity of the $T_{p, p'}$- and $\widehat{T}_{p, p'}$-family fails in general. Nevertheless, one can still diagonal the matrices $T_{p, p'} (q), \widehat{T}_{p, p'}(q)$ by using two invertible $q$-independent matrices $P_B, P_W$  so that both $P_W^{-1} T_{p, p'} (q)P_B, P_B^{-1} \widehat{T}_{p, p'}(q)P_W$ are diagonalized (\cite{BBP} (2.34)). For convenience, we shall refer those diagonal entries as the "eigenvalues" of $T_{p, p'}, \widehat{T}_{p, p'}$.  By (\req(TTc)), the diagonal forms of $T_{p, p'}, \widehat{T}_{p, p'}$ are related by (\cite{BBP} (4.46)):
\be
\widehat{T}_{p, p'}(q)  = (\frac{ f_{pq}}{ f_{p'q}})^L T_{p, p'} (q) D 
\ele(TTD)
where $D$ is a $q$-independent diagonal matrix. 

The CPM transfer matrices in (\req(Tpq)) are the $Q_R, Q_L$-operator of the $\tau^{(2)}$-matrix, satisfying the $\tau^{(2)}T$-relations (\cite{BBP} (4.20) (4.21), \cite{R0710}):
\bea(l)
\tau^{(2)}(t_q) T_{p, p'}(Uq) = \{\frac{(y_p-  \omega x_q)(t_{p'}- t_q) }{y_p y_{p'}(x_{p'}-  x_q)}\}^L T_{p ,p'}( q) + \{\frac{(y_{p'}- y_q)(t_p- \omega t_q) }{y_p y_{p'}(x_p-y_q)}\}^L T_{p, p'}(R^2  q) , \\

\widehat{T}_{p, p'}(U q) \tau^{(2)}(t_q)  = \{\frac{(y_{p'}- \omega x_q)(t_p- t_q) }{y_p y_{p'}(x_p- x_q)}\}^L \widehat{T}_{p, p'}( q) + \{\frac{(y_p- y_q)(t_{p'}- \omega t_q) }{y_p y_{p'}(x_{p'}-y_q)}\}^L \widehat{T}_{p, p'}(R^2 q) , 
\elea(tauT)
where $U, R$ are in (\req(Aut)). Using (\req(Tqrel)),  one can write the first $\tau^{(2)}T$-relation in (\req(tauT)) in an equivalent form in terms of automorphisms $U$ or $U'$ (\cite{BBP} (4.31)): 
\bea(l)
\tau^{(2)}(t_q) T_{p, p'}(Uq) = \varphi_q T_{p ,p'}( q) + \overline{\varphi}_{Uq} X T_{p, p'}(U^2 q) ,  \\
\tau^{(2)}(t_q) T_{p, p'}(U'q) = \varphi_q^\prime  X T_{p, p'}(q) + \overline{\varphi}_{U'q}^\prime T_{p, p'}(U'^2 q) , 
\elea(tauTU)
where 
\bea(ll)
\varphi_q  (= \varphi_{p, p'; q}) = \{\frac{(t_{p'}- t_q)  (y_p-  \omega x_q)}{y_p y_{p'}(x_{p'}-  x_q)}\}^L, & \overline{\varphi}_q (=\overline{\varphi}_{p, p';q}) =
\{\frac{\omega \mu_{p'} \mu_p(t_p- t_q)(x_{p'}- x_q) }{y_p y_{p'}(y_p- \omega x_q)}\}^L , \\
\varphi_q^\prime  (= \varphi_{p, p';q}^\prime ) =  
\{\frac{\omega \mu_p \mu_{p'}(t_{p'}- t_q)(x_p- y_q) }{y_p y_{p'}(y_{p'}- y_q)}\}^L, &  
\overline{\varphi}_q^\prime  (=\overline{\varphi}_{p, p'; q}^\prime ) = \{\frac{(t_p- t_q)  (y_{p'}-  y_q)}{y_p y_{p'}(x_p -  y_q)}\}^L .
\elea(varphi)
Similarly, the second $\tau^{(2)}T$-relation in (\req(tauT)) can be written equivalently as 
\bea(l)
\widehat{T}_{p, p'}(U q) \tau^{(2)}(t_q)  = \varphi_{p',p; q} \widehat{T}_{p, p'}( q) + \overline{\varphi}_{p', p;Uq}  X \widehat{T}_{p, p'}(U^2 q) , \\
\widehat{T}_{p, p'}( U' q) \tau^{(2)}(t_q)  = \varphi_{p', p; q}^\prime  X \widehat{T}_{p, p'}(q) + \overline{\varphi}_{p', p; U'q}^\prime  \widehat{T}_{p, p'}(U'^{2} y_q) .
\elea(tauTh)
Using $\tau^{(2)}T$-relation  (\req(tauTU)) and the recursive fusion relation (\req(fus)), one finds $\tau^{(j)}T$-relation (\cite{BBP} $(4.34)_{k=0}$):
\bea(ll)
\tau^{(j)}(t_q)= &\sum_{m=0}^{j-1} \varphi_q \varphi_{Uq} \cdots \varphi_{U^{m-1}q}
\overline{\varphi}_{U^{m+1}q} \overline{\varphi}_{U^{m+2}q}  \cdots \overline{\varphi}_{U^{j-1}q} \\
&\times T_{p, p'}(x_q, y_q)T_{p, p'}(\omega^m  x_q, y_q)^{-1} T_{p, p'}(\omega^j x_q, y_q) T_{p, p'}(\omega^{m+1} x_q,  y_q)^{-1}X^{j-m-1} .
\elea(tjT)

\section{Bethe equation of $\tau^{(2)}$-model \label{sec:Bethe}}
\setcounter{equation}{0}
In this section, we establish the Bethe equation of $\tau^{(2)}$-model by the Wiener-Hopf splitting method in \cite{MR}.   
\subsection{ A general description of $\tau^{(2)}$-eigenvalues \label{ssec.t2TB}}
First we describe a mechanism of constructing $\tau^{(2)}(t)$  as a general solution of equation (\ref{tau2eq}). Let $s$ be a (system of) variable (or variables) algebraically dependent on $t$, $s_0$ a base point corresponding to $t=0$, and  $s \mapsto \phi (s)$ an order-$N$ automorphism so that $t$ can be expressed as a regular function of $s$ with  $\phi (s)$ corresponding to $\omega t$.
Assume $z(t)$ in (\req(zCP)) can be factored as the product of two $s$-functions, ${\it H}^+(s)$ and ${\it H}^- (s)$, with respect to $\alpha_q, \overline{\alpha}_q$ in (\req(aaq)), such that the following relations hold:
\bea(ll)
z(t) = {\it H}^+(s) {\it H}^-(s), ~ ~ ~ ~  H^+(s_0) = 1 ,& \\
\alpha_q = {\it H}^+(s) {\it H}^+(\phi (s)) \cdots {\it H}^+(\phi^{N-1}(s)), & \overline{\alpha}_q = {\it H}^-(s) {\it H}^-(\phi (s)) \cdots {\it H}^-(\phi^{N-1}(s)). 
\elea(cHpm) 
The above relations are consistent with (\req(aaz)). Note that (\req(cHpm)) remains valid when replacing ${\it H}^\pm (s)$ by $\gamma^{\pm 1} {\it H}^\pm (s)$  with $\gamma^N = 1$. 
Using ${\it H}^\pm (s)$ in (\req(cHpm)), one can verify that an exact solution of (\ref{tau2eq}) is given by
\be
 \tau^{(2)}(t) = {\it H}^+(s) \frac{\Lambda (s )}{\Lambda( \phi (s))} + \omega^Q {\it H}^-(\phi (s))\frac{\Lambda(\phi^2(s))}{\Lambda( \phi (s))}, 
\ele(t2sl)
where $\Lambda(s )$ is a function with the constraint that the function in above right-hand side defines a $t$-function, (a condition automatically true in case $s = t$ and $\phi (s ) = \omega t$). Using (\req(t2sl)) and the recursive fusion relation in (\req(fus)), one finds the following expression of $\tau^{(j)}$-eigenvalue: 
\bea(rl)
\tau^{(j)}(t)&= \omega^{jQ} \Lambda(s ) \Lambda(\phi^j(s)) \sum_{m=0}^{j-1} \bigg ( {\it H}^+(s) {\it H}^+(\phi(s))\cdots {\it H}^+(\phi^{m-1}s)  \\
\times& {\it H}^-(\phi^{m+1}(s)) {\it H}^-(\phi^{m+2}(s)) \cdots {\it H}^-(\phi^{j-1}(s)) \Lambda(\phi^m(s))^{-1}  \Lambda(\phi^{m+1}(s))^{-1} \omega^{-(m+1)Q} \bigg).
\elea(tjF)
In particular,  by setting $s = (x_q, y_q, \mu_q)$ for $q$ being the rapidity in (\req(xymu)) or (\req(Cdeg)),  and $s_0$ being the element whose $x$-value equals to 0, the relations, (\req(zCP)) and (\req(aaq)), imply (\req(cHpm)) holds for $({\it H}^+(s), {\it H}^-(s)) = (\varphi_q, \overline{\varphi}_q) $ with $\varphi_q, \overline{\varphi}_q$ in (\req(varphi)). The formula (\req(t2sl)) (with $\phi$ = $U$ in (\req(Aut)))  is an equivalent form of $\tau^{(2)}T$-relation (\req(tauTU)), where $\Lambda(s )$ is a $T_{p, p'}$-eigenvalue. The relation (\req(tjF)) is equivalent to $\tau^{(j)}T$-relation (\req(tjT)). Similarly, the first relation in (\req(tauTh)) is the same as (\req(t2sl)) with $({\it H}^+(s), {\it H}^-(s)) = (\varphi_{p', p; q}, \overline{\varphi}_{p'p; q})$ and $\phi$ = $U$. One may consider another pair of functions, ${\it H}'^-(s)$ and ${\it H}'^+(s)$, an order $N$ automorphism $\phi'$, and a base point $s'_0$ with the relation
\bea(ll)
z(t) = {\it H}'^-(s) {\it H}'^+(s), ~ ~ ~ ~ H'^+(s'_0) = 1 , &\\
\alpha_q = {\it H}'^-(s) {\it H}'^-(\phi' (s)) \cdots {\it H}'^-(\phi'^{N-1}(s)), &
\overline{\alpha}_q = {\it H}'^+(s) {\it H}'^+(\phi' (s)) \cdots {\it H}'^+(\phi'^{N-1}(s)).  
\elea(H'pm)
Write  the solutions of (\ref{tau2eq}) in the form with the factor $\omega^Q$ in front:
\bea(l)
\tau^{(2)}(t) = \omega^Q {\it H'}^- (s) \frac{ \Lambda'(s )}{\Lambda'( \phi' (s))} + {\it H'}^+ (\phi (s))\frac{\Lambda'(\phi^2(s))}{\Lambda'( \phi' (s))} .
\elea(t2wQ)
By setting $({\it H'}^-(s), {\it H'}^+(s))$ = $(\varphi'_q, \overline{\varphi}'_q)$ or $(\varphi'_{p', p; q}, \overline{\varphi}'_{p'p; q})$, one finds the other  $\tau^{(2)}T$-relation in (\req(tauTU)) or (\req(tauTh)) with $s'_0$ the element with zero $y$-coordinate, $\phi'= U'$, and $\Lambda(s )$ being an eigenvalue of $T_{p, p'}$ or $\widehat{T}_{p, p'}$.
Note that  one may interchange  the functions $H^\pm$ in (\req(t2sl)), or $H'^\pm$ in (\req(t2wQ)), to form $\tau^{(2)}(t)$. The resulting ones can be obtained from the previous ones by changing $Q, \tau^2(t)$ to $-Q, \omega^{-Q} \tau^2(t)$ and $\phi^{-1}, \Lambda (\phi^2(s))$  to $\phi, \Lambda(s)$. However, in order to identify $Q$ with charge of the eigenvalue $\tau^{(2)}(t)$, we need only to consider the cases (\req(t2sl)) and (\req(t2wQ)).

\subsection{ Bethe equation of chiral Potts model with alternating vertical rapidities  \label{ssec.Bethe}}
In solving the eigenvalue problem of the $\tau^{(2)}$-matrix, 
it will be convenient to choose the variable $s= t$ with $\phi ( t ) = \phi' ( t )= \omega t$ in (\req(t2sl)) and (\req(t2wQ)). In this section, we follow the method in \cite{MR} to construct the $t$-functions $h^\pm (t)$ as $H^\pm$ and ${h'}^\mp (t)$ as ${\it H'}^\mp$  through the Wiener-Hopf splitting of $\alpha_q, \overline{\alpha}_q$. With $\Lambda(s)$ defined by
\bea(l)
\Lambda(t) = t^{P_a} F(t) , \ \ F (t) = \prod_{j=1}^J (1+ \omega v_j t)  ~ ~ \ \ (v_j^N \neq v_i^N \ {\rm for} ~ j \neq i), 
\elea(Fpol)
where $P_a$ is an integer, (\req(t2sl))  is expressed by
\be
\tau^{(2)}(t) = \omega^{-P_a} h^+(t) \frac{ F (t)}{F( \omega t)} + \omega^{Q+P_a} h^-(\omega t)\frac{F(\omega^2 t)}{F( \omega t)} . 
\ele(t2F)
The regular-function condition of $\tau^{(2)}(t)$ 
requires the roots of $F (t)$ satisfy the Bethe equation:
\be
\prod_{j=1}^J \frac{v_i -  \omega^{-1}   v_j }{ v_i -\omega  v_j } = - \omega^{Q+2P_a} \frac{h^-(-  \omega^{-1} v_i^{-1})}{h^+(-  \omega^{-2} v_i^{-1} )}, \ \ i= 1, \ldots, J.
\ele(Bethet2)
By (\req(tjF)), the functions $\tau^{(j)}(t) ~ (j \geq 2 )$  are expressed by 
\bea(l)
\tau^{(j)}(t)= \omega^{(j-1)(Q+P_a)} F(t ) F(\omega^j t ) \sum_{n=0}^{j-1} \frac{h^+(t) \cdots h^+(\omega^{n-1} t) 
h^-( \omega^{n+1}t) \cdots h^-(\omega^{j-1} t)\omega^{-n(Q+2P_a)} }{ 
 F(\omega^n t)  F(\omega^{n+1}t)} ; \\
\tau^{(N)}(t)=\omega^{-Q-P_a}  F(t )^2 \sum_{n=0}^{N-1} \frac{h^+(t) \cdots h^+(\omega^{n-1} t) 
h^-( \omega^{n+1}t) \cdots h^-(\omega^{N-1} t)\omega^{-n(Q+2P_a)} }{ 
 F(\omega^n t)  F(\omega^{n+1}t)}. 
\elea(tjFt)
The Bethe equation (\req(Bethet2)) can also be characterized as the regular-function criterion of any one of the above $\tau^{(j)}(t)$s. Similarly, when using the expression  (\ref{t2wQ}) with $H'^\mp (s) = h'^\mp (t)$, and $\Lambda'(s) = t^{P_b} F'(t)$ with $F' (t) = \prod_{j=1}^{J'} (1+ \omega v'_j t)$ as in (\req(Fpol)),
\bea(l)
\tau^{(2)}(t) = \omega^{Q-P_b}  h'^-(t) \frac{ F' (t)}{F'( \omega t)} + \omega^{P_b} h'^+ (\omega t)\frac{ F'(\omega^2 t)}{F'( \omega t)}, 
\elea(t2F')
the Bethe equation takes the form
\be
\prod_{j=1}^{J'} \frac{v'_i -  \omega^{-1}   v'_j }{ v'_i -\omega  v'_j } = - \omega^{-Q+2P_b} \frac{h'^+(-  \omega^{-1} v'^{-1}_i)}{h'^-(-  \omega^{-2} v'^{-1}_i )}, \ \ i= 1, \ldots, J',
\ele(Bethet2')
with the $\tau^{(j)}$-expression 
\bea(l)
\tau^{(j)}(t)= \omega^{(j-1)P_b} F'(t ) F'(\omega^j t ) \sum_{n=0}^{j-1} \frac{h'^-(t) \cdots h'^-(\omega^{n-1} t) h'^+( \omega^{n+1}t) \cdots h'^+(\omega^{j-1} t)\omega^{n(Q-2P_b)} }{ 
 F'(\omega^n t)  F'(\omega^{n+1}t)} ; \\
\tau^{(N)}(t)=\omega^{-P_b}  F'(t )^2 \sum_{n=0}^{N-1} \frac{h'^-(t) \cdots h'^-(\omega^{n-1} t) h'^+( \omega^{n+1}t) \cdots h'^+(\omega^{N-1} t)\omega^{n(Q-2P_b)} }{ 
 F'(\omega^n t)  F'(\omega^{n+1}t)}. 
\elea(tjFt')

Next we construct the functions $h^\pm (t), h'^\mp (t)$ in the Bethe equation (\req(Bethet2)) , (\req(Bethet2')) for   a  rapidity curve in (\req(xymu)) or (\req(Cdeg)). 
Write the functions in (\req(aaq)) in the form
\bea(l)
\alpha_q = ( \frac{t_{p'}^N-t^N}{y_p^N y_{p'}^N})^{L} \beta (\sigma_q)^L  , \ \ \overline{\alpha}_q   = (\frac{\mu_p^N \mu_{p'}^N (t_p^N-t^N)}{y_p^N y_{p'}^N })^L 
\beta (\sigma_q)^{-L} 
\elea(abF)
where $\beta (\sigma)$ is the following $\sigma$-function for the variable $\sigma$ in (\req(sigma)): 
\bea(lll)
\beta (\sigma) =  \frac{1- \lambda_p \lambda }{1-\lambda_{p'}^{-1} \lambda } ~ ~   {\rm for} \ W_{k'} ~ {\rm and}   ~ W_1^\prime ; &
 \frac{ 1-{\bf x}_p - {\bf x}}{ {\bf x}_{p'}-{\bf x}} ~ ~ {\rm for} \  W_1^{\prime \prime} ; &
\frac{- {\bf x}_p - {\bf x}  }{{\bf x}_{p'}- {\bf x}} ~ ~  {\rm for} \ W_1^{\prime \prime \prime} .
\elea(beta)
Let $C_t$ be a contour in the $t$-plane circled clock-wise around the interval between the elements, $t_0^+$ and $t_0^-$, defined in (\req(t0)). With $t$ outside the contour, we define the function
\be
e^+ (t) = {\rm exp} \bigg[  \frac{L}{2 N \pi {\rm i}} \int_{C_t} \frac{dt'}{t'-t} \ln \beta( \sigma') \bigg] 
\ele(ef)
where $\sigma, \sigma'$ are in the "{\it positive}" sheet $W^+$ of $W$ in (\req(hW)): $|\lambda| < 1$ for $W_{k'}$ and $W_1^\prime$; $\Re ({\bf x}) < \frac{1}{2} $ for $W_1^{\prime \prime}$, and $\Re ({\bf x}) < 0$ for $W_1^{\prime \prime \prime}$. 
Note that outside the contour $C_t$, the $t$-domain can be lifted to the hyperelliptic curve $W$ with the assigned $\sigma$-value, by which $\sigma$ can be regarded as a function of $t$, hence $e^+ (t)$ coincides with the algebraic-function 
\be
\beta (\sigma)^\frac{L}{N} = (\frac{y_p^N - x^N}{x_{p'}^N- x^N})^\frac{L}{N} = \omega^L \mu_p^L \mu_{p'}^L (\frac{x_p^N - y^N}{y_{p'}^N- y^N})^\frac{L}{N}. 
\ele(ealg)
More precisely, $e^+(t)$ is expressed by (\req(ealg)) with the values of $x$ and $y$ restricted on the positive region $W^+$ of $W$ in (\req(hW)):
\bea(lllll)
W^+ := & W_{k'}^+ : & | 1 - k x^N | \geq |k'|, | 1 - k y^N | \leq |k'|    ; &
 W_1^{\prime +}:& | 1 - x^N | \geq 1 , | 1 -  y^N | \leq 1  ; \\
& W_1^{\prime \prime +}:& \Re ( 1 - x^N ) = \Re ( y^N ) \geq \frac{1}{2}  ; &
W_1^{\prime \prime \prime +}:& - \Re (x^N ) = \Re (y^N ) \geq 0   . \\
\elea(Dom)
In particular, when $t=0$, one takes $x =0$ with the corresponding $\sigma$-value being $k', 1 , 1, 0 $ respectively. For later use, we denote the negative region $W^-$ of $W$ by  reversing inequality signs in (\req(Dom)). Let $e^- (t)$ be the function defined by formula (\req(ef)) where $\sigma'$  is in the {\it negative} sheet $W^-$ of $W$,  hence $e^- (t)$ again expressed by (\req(ealg)). 
We may assume $e^\pm ( \omega^{N} t ) = e^\pm ( t )$ by a suitable choice of phase-factors. Note that $e^\pm (t)$ are indeed functions defined on the region $W^\pm$, and we shall also write
$$
e^\pm (t) = e^\pm (t, \sigma) \ ~ ~  {\rm for} \ (t, \sigma) \in W^\pm ,
$$  
whenever a clearer expression will be needed.

Using the function in (\req(ef)), we define the function $h^\pm (t)$ in (\req(t2F)) through the Wiener-Hopf splitting of $\alpha_q, \overline{\alpha}_q$ as formula (24a,b) in \cite{MR}:
\be 
h^+(t) =  [\frac{t_{p'}-t}{y_p y_{p'}}]^L e^+ (t) \gamma, ~ ~ h^-(t) =  [ \frac{\omega \mu_p \mu_{p'}(t_p -t)}{y_p y_{p'}} ]^L \frac{1}{e^+(t) \gamma}  
\ele(hpm) 
where $\gamma$ is a constant factor determined by the condition $h^+(0)=1$. Indeed, the evaluation of the integral in (\req(ef)) shows that $\gamma$ is a $N$th root of unity, 
$\gamma = \omega^{P_\gamma}$, 
with 
\bea(ll)
\lim_{t \rightarrow \infty} \frac{ h^+ (t)}{t^L}  = \gamma (\frac{-1}{y_p y_{p'}})^L, & \lim_{t \rightarrow \infty} \frac{ h^- (t)}{t^L}  = \gamma^{-1} (\frac{-\omega \mu_p \mu_{p'}}{y_p y_{p'}})^L . 
\elea(hpml)
The expression (\req(hpm)) yields the condition (\req(cHpm)) with $H^\pm = h^\pm (t)$:
\bea(llll)
z(t) = h^+(t) h^-(t), &
\alpha_q = \prod_{j=0}^{N-1} h^+(\omega^j t) , & \overline{\alpha}_q = \prod_{j=0}^{N-1} h^-(\omega^j t), & h^+(0) = 1 . 
\elea(hpd) 
Hence we can express  $\tau^{(j)}(t)$ by (\req(tjFt)) using $h^\pm(t)$ in (\req(hpm)) and the Bethe solution $F(t)$ for (\req(Bethet2)). One may use $e^-(t)$ to define the Wiener-Hopf splitting $h'^\mp$ of $\alpha_q, \overline{\alpha}_q$: 
\be 
h'^-(t) =  [\frac{t_{p'}-t}{y_p y_{p'}}]^L e^- (t) \gamma', ~ ~ h'^+(t) =  [ \frac{\omega \mu_p \mu_{p'}(t_p -t)}{y_p y_{p'}} ]^L \frac{1}{e^-(t) \gamma'}  
\ele(h'pm) 
where  $\gamma' = \omega^{P_{\gamma'}}$ is determined by the condition $h'^+(0)=1$. Then one has
\bea(ll)
\lim_{t \rightarrow \infty} \frac{ h'^- (t)}{t^L}  = \gamma' (\frac{-1}{y_p y_{p'}})^L ( \mu_p^N \mu_{p'}^N)^\frac{L}{N} , & \lim_{t \rightarrow \infty} \frac{ h'^+ (t)}{t^L}  = \gamma'^{-1} (\frac{-\omega \mu_p \mu_{p'}}{y_p y_{p'}})^L
( \mu_p^N \mu_{p'}^N)^\frac{-L}{N} . 
\elea(h'pml)
Using (\req(ealg)), one finds the algebraic-function-expression of $h^\pm, h'^\mp$:
\bea(l) 
( \frac{(t_{p'}-t)^L}{y_p^L y_{p'}^L} (\frac{y_p^N - x^N}{x_{p'}^N- x^N})^\frac{L}{N} ,    \frac{(t_p -t)^L}{y_p^L y_{p'}^L}    (\frac{y_{p'}^N- y^N}{x_p^N - y^N})^\frac{L}{N} ) = \left\{ \begin{array}{ll} ( \gamma^{-1} h^+(t) , \gamma h^-(t) ) & {\rm for} \ \ (t, \sigma ) \in W^+ ;  \\
( \gamma'^{-1} h'^-(t) , \gamma' h'^+(t) ) & {\rm for} \ \ (t, \sigma ) \in W^-.  \end{array} \right.
\elea(hh'AF) 
Note that for the general $p$ and $p'$, the above expression holds {\it only}  on a "half" of  Riemann surface $W$, i.e. $W^+$ (or $W^-$) where $t$ can be considered as an algebraic function of $\sigma$ in the restricted domain, but not on the entire $W$ except the special alternating superintegrable case discussed later in section \ref{ssec.AlSup}. Accordingly, the relation (\req(t2F)) (or (\req(t2F'))) is valid by regarding $t$ as a function of the variable $\sigma$, and it becomes the $t$-polynomial relation  only in the alternating superintegrable case.  The relation (\req(H'pm)) holds with $H'^\mp = h'^\mp (t)$:
\bea(llll)
z(t) = h'^-(t) h'^+(t), &
\alpha_q = \prod_{j=0}^{N-1} h'^-(\omega^j t) , & \overline{\alpha}_q = \prod_{j=0}^{N-1} h'^+(\omega^j t), & h'^+(0) = 1 ,
\elea(h'pd) 
with the $\tau^{(j)}(t)$-expression  (\req(tjFt')) by using $h'^\mp(t)$in (\req(h'pml)) and  the Bethe solution $F'(t)$ of (\req(Bethet2')). 
\par \noindent 
{\bf Remark}: The Wiener-Hopf splittings, (\req(hpd)) and (\req(h'pd)),  of $\alpha_q, \overline{\alpha}_q$ are related as follows.  In the previous discussion, it is understood that the $t$-functions, $h^\pm (t)$ and $h'^\mp (t)$, are indeed functions on one sheet of  the hyperelliptic curve $W$:  
$$
\begin{array}{l}
h^\pm (t) = h^\pm ( t, \sigma ), \ \tau^{(j)}(t) = \tau^{(j)}(t, \sigma) \ \ \ \ ( t, \sigma) \in W^+ ; \\
h'^\mp (t) = h'^\mp ( t, \sigma ), \ \tau^{(j)}(t) = \tau^{(j)}(t, \sigma) \ \ \ \ ( t, \sigma) \in W^- . 
\end{array}
$$
For convenience, we define $e^+(t)^*$ to be the function on $W^+$ by changing $x^N$ in $e^+(t)$ to $y^N$:
$$
e^+(t)^* ( = e^+(t, \sigma)^*) = (\frac{y_p^N - y^N}{x_{p'}^N- y^N})^\frac{L}{N}, \ \ \ \ (t, \sigma) \in W^+ ,
$$
and  $h^\pm (t)^*$,  $\tau^{(j)}(t)^*$ are the functions of $W^+$ by replacing $e^+(t)$ by $e^+(t)^*$ in (\req(hpm)), (\req(tjF)) respectively. Using (\req(ealg)), one finds
$$
e^+(t)^*  = e^+(t, \sigma)(\frac{\overline{\alpha}_q}{\alpha_q })^\frac{1}{N} = e^+(t)(\frac{\overline{\alpha}_q}{\alpha_q })^\frac{1}{N}  ~ ~ (= e^- ( t, \sigma^\dagger )) = e^-(t)\ \ \ {\rm for} \ ( t, \sigma) \in W^+. 
$$
Hence one obtains
$$
h^+ (t)^* = h^+ (t) (\frac{\overline{\alpha}_q}{\alpha_q })^\frac{1}{N} = \gamma \gamma'^{-1}  h'^- (t), \ \ h^- (t)^* = h^- (t) (\frac{\overline{\alpha}_q}{\alpha_q })^\frac{-1}{N} = \gamma'\gamma^{-1}  h'^+ (t).  
$$ 
For generic $p$ and $p'$, $\tau^{(j)}(t)^*$ is not a regular function, hence not equal to the $t$-polynomial $\tau^{(j)}(t)$. Indeed by (\req(tjFt)),  $\tau^{(j)}(t)^* , (j \geq 2),$ is expressed by 
$$
\begin{array}{l}
\tau^{(j)}(t)^* = \omega^{(j-1)(Q+P_a)}  F(t)F(\omega^j t) \sum_{n=0}^{j-1} \frac{h^+(t)^* \cdots h^+(\omega^{n-1} t)^* h^-( \omega^{n+1}t)^* \cdots h^-(\omega^{j-1} t)^* \omega^{-n(Q+2P_a)} }{ F(\omega^n t)  F(\omega^{n+1}t)} \\ 
=\omega^{(j-1)(Q+P_a)}  F(t ) F(\omega^j t) \sum_{n=0}^{j-1} \frac{h^+(t) \cdots h^+(\omega^{n-1} t) h^-( \omega^{n+1}t) \cdots h^-(\omega^{j-1} t) (\omega^{-(Q+2P_a)}(\frac{\overline{\alpha}_q}{\alpha_q })^\frac{2}{N})^n }{ F(\omega^n t)  F(\omega^{n+1}t)} (\frac{\overline{\alpha}_q}{\alpha_q })^\frac{-j+1}{N}. 
\end{array}
$$
By (\req(Bethet2)), the condition of $\tau^{(j)}(t)^*$ being regular at zeros of $F(\omega^{n+1} t)$ for $0 \leq n \leq j-2$ is the requirement of $(\frac{\overline{\alpha}_q}{\alpha_q })^\frac{2}{N} =1$ when $t= - \omega^{-1}v_i^{-1} ~ (1 \leq j \leq J)$. If the functions $\alpha_q $ and $\overline{\alpha}_q $ are not equal,  $\tau^{(j)}(t)^*$ is not finite for $t= t_p, t_{p'}$. Using the relation between $h^\pm(t)^*$ and $h'^\pm(t)$, one concludes that the polynomials $F(t), F'(t)$ in (\req(tjFt)), (\req(tjFt')) are not the same when $\alpha_q \neq \overline{\alpha}_q$. Indeed by (\req(Bethet2)) and (\req(Bethet2')),  $F(t)= F'(t)$ is provided by $\frac{h^-(\omega t)^N}{h^+(t )^N}  =  \frac{h'^+(\omega t)^N}{h'^-(t )^N}$ , equivalent to $\overline{\alpha}_q = \pm \alpha_q $, which is the same conclusion for the polynomial condition of $\tau^{(j)}(t)^*$. In particular, when $\overline{\alpha}_q = \alpha_q $, i.e. the alternating superintegrable case in section \ref{ssec.AlSup}, by (\req(tjFt')), one finds $\tau^{(j)}(t)^*= \tau^{(j)}(t)$ with $P_b \equiv Q+P_a+P_{\gamma'}-P_\gamma$.

\section{Eigenvalues of Chiral Potts Transfer Matrix with Alternating Rapidities \label{sec:EiCH}}
\setcounter{equation}{0}
In this section, we discuss the eigenvalue problem of the transfer matrices $T_{p,p'}, \widehat{T}_{p, p'} $ in (\req(Tpq)) for a rapidity curve  in (\req(xymu)) or (\req(Cdeg)) by employing the functions $h^\pm (t), h'^\mp (t) $ in (\req(hpm)) (\req(h'pm)). As in the homogeneous CPM case \cite{B90}, we normalize the transfer matrices by multiplying a power of $g_p(q)$ and $\overline{g}_p (q)$ in (\req(gg1)) to eliminate of the factor appeared in formula (\req(Tqrel)). Note that  $\mu_q^\frac{N(N-1)}{2} g_p(q)$ is a function of $x_q$ and $y_q$ only, independent of $\mu_q$, whose value under the automorphism $T$ in (\req(Aut)) changes by  
\be
\mu_{Tq}^\frac{N(N-1)}{2}  g_p(Tq) = \frac{  x_p^N -  y_q^N }{ y_p^N  -x_q^N } (\frac{   y_p - \omega x_q }{  \omega x_p - y_q })^N \mu_q^\frac{N(N-1)}{2}  g_p(q) .
\ele(gT)
The function $\overline{g}_p (q)$ also depends on $x_q$ and $y_q$ only, i.e.  $\overline{g}_p (q)= \overline{g}_p (\overline{q})$ with $(x_{\overline{q}}, y_{\overline{q}}, \mu_{\overline{q}})= (x_q, y_q,, \omega \mu_q)$, which follows from $\overline{W}^f_{p, \overline{q}}(0)= \overline{W}^f_{p, q}(1)$ and $\frac{\overline{W}^f_{p \overline{q}}(n)}{\overline{W}^f_{p \overline{q}}(0)}= \frac{\overline{W}^f_{p, q}(n+1)}{\overline{W}^f_{p, q}(1)}$. By (\req(gg1)), The relations, $\overline{W}^f_{p ~ Tq}(0)=   \frac{ -(y_p - \omega^{-1} y_q) }{\mu_p \mu_q(x_p - x_q)} \overline{W}^f_{p q}(0)$ and $(\mu_p \mu_q)^N (x_p^N- x_q^N)= y_q^N- y_p^N $, in turn yield 
\be
\overline{g}_p (Tq) =  \frac{x_p^N- x_q^N }{ y_p^N - y_q^N } (\frac{ y_p - \omega^{-1} y_q }{x_p - x_q})^N \overline{g}_p (q). 
\ele(g-T)
By $\mu_p^N  \mu_{p'}^N (x_p^N-y_q^N)(x_{p'}^N-x_q^N) =(y_p^N-x_q^N)(y_{p'}^N-y_q^N)$, the relations, (\req(gT)) and (\req(g-T)), imply 
\be
 \mu_{Tq}^\frac{N(N-1)}{2} g_p(Tq)\overline{g}_{p'}(Tq) =   (\frac{ ( y_p- \omega x_q ) (y_{p'} - \omega^{-1} y_q)}{\mu_p  \mu_{p'} (\omega x_p -  y_q) (x_{p'} -  x_q)  })^N  \mu_q^\frac{N(N-1)}{2} g_p(q)\overline{g}_{p'}(q) \omega^{N(N+1)} \ .
\ele(gg)
Comparing the factors in (\req(Tqrel)) and (\req(gg)), we consider the normalized transfer matrices as in \cite{B90} Sect. 2 :
\bea(l)
V (q) ~ ( = V (x_q, y_q)) :=  T_{p, p'} (q) \bigg(  \mu_q^\frac{N(N-1)}{2} g_p(q) \overline{g}_{p'}(q) \bigg)^{\frac{-L}{N}} ,  \\
\widehat{V} (q) ~ ( = \widehat{V} (x_q, y_q)) :=  \widehat{T}_{p, p'} (q) \bigg(  \mu_q^\frac{N(N-1)}{2} g_{p'} (q) \overline{g}_p (q) \bigg)^{\frac{-L}{N}} .
\elea(Vdef)
By (\req(TTD)), the eigenvalues of $V (q)$ and $\widehat{V} (q)$ are related by
\be
\widehat{V}(q) = V (q) D
\ele(VVh)
where $D$ is a $q$-independent scale. By (\req(Tqrel)), the following relations hold (see, \cite{B90} page 111):  
\be
V (\omega x_q, \omega^{-1} y_q )= \omega^{-L}X^{-1} V (x_q, y_q) , ~ ~ ~ \widehat{V} (\omega x_q, \omega^{-1} y_q )=   \omega^{-L} X^{-1} \widehat{V} (x_q, y_q) ,
\ele(VVT)
by which $V (q), \widehat{V} (q)$ depend on the values of $(t_q, \sigma_q)$ in (\req(sigma)), up to a $N$th root of unity\footnote{More precisely, $V (t_q, \sigma_q ), \widehat{V}(t_q, \sigma_q )$ are indeed  meromorphic sections of a $N$th torsion line bundle  associated to the $N$-fold covering $x y$-curve over $W$ in (\req(hW)). }. In particular, $V (x_q, y_q)^N$ is a meromorphic functions of $(t_q, \sigma_q)$.

\subsection{Functional relations of the chiral Potts model  \label{ssec.FnRel}}
The transfer chiral Potts matrix and the $\tau^{(j)}$-matrix for rapidities in (\req(xymu)) are known to satisfy a set of functional relations \cite{BBP}, which again hold in the degenerate models with rapidities in (\req(Cdeg)) (\cite{R0710} section 5). Among the functional relations are the fusion relation (\req(fus)), $\tau^{(2)}T$-relation (\req(tauTU)), $\tau^{(j)}T$-relation (\req(tjT)), and the $T\hat{T}$-relation (\cite{B02} (13), \cite{R05o} (15)):
\be
 \frac{T_{p, p'}(x_q, y_q) \widehat{T}_{p, p'}(y_q, \omega^j x_q)}{r_{p'}(q) h_{j; p, p'}(q) } = \tau^{(j)} (t_q) + \frac{z(t_q)z(\omega t_q) \cdots z(\omega^{j-1} t_q)}{\alpha_q } \tau^{(N-j)} (\omega^j t_q) X^j  
\ele(ThatT)
where  $r_{p'}(q)= (\frac{N(x_{p'}-x_q)(y_{p'}-y_q)(t_{p'}^N -t_q^N)}{(x_{p'}^N- x_q^N) (y_{p'}^N -y_q^N)(t_{p'}-t_q)})^L $, $h_{j; p, p'}(q)=( \prod_{m=1}^{j-1} \frac{y_p y_{p'} (x_{p'}- \omega^m x_q)}{(y_p-\omega^m x_q)(t_{p'}-\omega^m t_q)} )^L $,  $z(t)$ and $\alpha_q$ in (\req(zCP)), (\req(aaq)) respectively.
By (\req(tjT)) and (\req(ThatT)), follows the functional relation of the transfer matrix (\cite{BBP}(4.40)):
$$
\widehat{T}_{p, p'}(y_q, x_q) = \sum_{m=0}^{N-1} C_m (q) T_{p, p'}(\omega^m x_q, y_q)^{-1}T_{p, p'}(x_q, y_q) T_{p, p'}(\omega^{m+1} x_q, y_q)^{-1} X^{-m-1}
$$
where $C_m (q) = \varphi(q) \cdots \varphi (U^{m-1}q)
\overline{\varphi} (U^{m+1}q)\cdots \overline{\varphi}(U^{N-1}q)( \frac{N(y_p y_{p'})^{N-1}(y_p-x_q)(y_{p'}-y_q)}{ (y_p^N - x_q^N) (y_{p'}^N - y_q^N)} )^L $.

For the eigenvalue problem of the chiral Potts transfer matrix, it is convenient to express the functional relations in terms of the normalized operators (\req(Vdef)).  By (\req(gg1)), (\req(g-q)) and
$$
\begin{array}{ll}
\frac{g_p (Uq)\overline{g}_{p'} (Uq)}{g_p (q) \overline{g}_{p'} (q)} = \frac{(y_p - \omega x_q)^N ( t_{p'} - t_q )^N (x_{p'}^N- x_q^N)}{ (x_{p'} - x_q)^N (y_p^N - x_q^N) (t_{p'}^N - t_q^N) }, &
\frac{g_p (U'q)\overline{g}_{p'} (U'q)}{g_p(q)\overline{g}_{p'} (q)} = \frac{ (x_p - y_q)^N ( t_{p'} - t_q )^N (y_{p'}^N- y_q^N)}{ (y_{p'} - y_q)^N (x_p^N- y_q^N) (t_{p'}^N - t_q^N) } ,
\end{array}
$$
the $\tau^{(2)}T$-relation (\req(tauTU))  takes the form (\cite{R04} (35)):
\bea(l)
\tau^{(2)}(t_q) V (Uq) =  (\frac{ (y_p^N - x_q^N) (t_{p'}^N - t_q^N) }  { y_p^N y_{p'}^N (x_{p'}^N- x_q^N)  })^\frac{L}{N} V ( q) 
+  z (\omega t_q)  (\frac{ y_p^N y_{p'}^N (x_{p'}^N- x_q^N)  }{ (y_p^N - x_q^N) (t_{p'}^N - t_q^N) }  )^\frac{L}{N}    X V (U^2 q).
\elea(tauV)
Note that one can also derive the above relation using (\req(tauTh)) and (\req(VVh)).

Denote 
\be
r_j (t_q, \sigma_q) := \alpha_q X^{-j}  \tau^{(j)} (t_q) + z(t_q)z(\omega t_q) \cdots z(\omega^{j-1} t_q) \tau^{(N-j)} (\omega^j t_q), ~ ~ 0 \leq j \leq N, 
\ele(rj)
where the variables $t, \sigma$ are in (\req(sigma)). By using (\req(Tqrel)), the $TT$-relation (\req(ThatT)) is the same as  
\be
\Gamma^{(j)}_q T_{p, p'}(x_q, y_q) \widehat{T}_{p, p'}(\omega^j y_q, x_q) = r_j (t_q, \sigma_q) 
\ele(GaTT)
where $\Gamma^{(j)}_q$ is defined by
$$
\Gamma^{(j)}_q = \{ \frac{\mu_p^j  \mu_{p'}^j (y_p^N-x_q^N)(y_{p'}^N -y_q^N)}{N  y_p^{N+j-1} y_{p'}^{N+j-1} (y_{p'}-y_q) (y_p - x_q)}  \prod_{m=0}^{j-1} (t_{p'}-\omega^m t_q)  \prod_{k=1}^j \frac{ \omega x_p - \omega^k y_q  }{  y_{p'}- \omega^k y_q } \}^L .
$$
Define
$$
\zeta (t) =  \zeta_0^\frac{-2L}{N} \prod_{k=1}^{N-1} z(\omega^k t)^\frac{k}{N}, ~ ~ \zeta_0 := e^\frac{ \pi {\rm i}(N-1)(N+4)}{12}.
$$
By (\req(gg1)) (\req(g-q)), the $TT$-relation (\req(GaTT)) becomes (\cite{B90} (8)):
\bea(ll)
\alpha_q^\frac{j}{N} \zeta (\omega^j t_q) V(x_q, y_q) \widehat{V} (\omega^j y_q, x_q) =  r_j (t_q, \sigma_q) , & 
\zeta ( t_q) V (x_q, y_q) \widehat{V}(y_q, x_q)  =   \tau^{(N)} (t_q) ,
\elea(VTT)
where the $\frac{j}{N}$th-power of $\alpha_q$ in (\req(aaq)) carries the phase-factor $\omega^{Lj}$. Hence we obtain
\bea(l)
 V^N (x_q, y_q) \prod_{j=1}^N \widehat{V} (\omega^j y_q, x_q) = \omega^\frac{-N(N+1)L}{2}\zeta_0^{2L} \alpha(\sigma_q) ^{-N}  \alpha (\sigma_q^\dagger)^\frac{-N+1}{2} \prod_{j=1}^N r_j (t_q, \sigma_q) ,  \\
(\prod_{j=1}^N V (\omega^j x_q, y_q)) (\prod_{j=1}^N \widehat{V} (y_q, \omega^j x_q ))= \zeta_0^{2L} \alpha (\sigma_q)^\frac{-N+1}{2} \alpha (\sigma_q^\dagger)^\frac{-N+1}{2} \prod_{j=1}^N \tau^{(N)} (\omega^j t_q) . 
\elea(ProdV)
By formula (3) of \cite{B90}, one finds  
$$
\prod_{j=0}^{N-1} T_{p, p'}(\omega^j x_q, y_q) = C s(y_q) \bigg(\prod_{j=1}^{N-1} (y_q - \omega^{j+1} x_p)^j(y_{p'} - \omega^j y_q)^j \bigg)^{-L} 
$$
for some $q$-independent constant $C$, and a matrix $s(y_q)$ of degree $N(N-1)L$ with $y_q$-rational-function entries, finite when both $x_q, y_q$ are finite. Hence
\be
\prod_{j=0}^{N-1} V(\omega^j x_q, y_q) = \zeta_0^L  \alpha (\sigma_q)^\frac{-N+1}{2} S(\sigma_q)
\ele(Vavg)
where the entries of matrix $S(\sigma_q)$ are  $\sigma_q$-rational functions with only possible pole at $\sigma =0$ ( \cite{B90} (4) )\footnote{For the rapidities in ${\goth W}_{k'}$ (\req(xymu)), $\sigma_q = \lambda_q$ by (\req(sigma)), and formula (\req(Vavg)) here is the same as \cite{B90} (4) except the factor$\lambda^{-(N-1)L/2}_q$. The difference is due to the extra factor $\mu_q^{N(N-1)/2}$ of $V(q)$ in (\req(Vdef)) we add in this paper.}. Using  (\req(VVh)), (\req(VVT)),  and (\req(VTT))-(\req(Vavg)), one can express $V (x_q, y_q)$ in terms of $S, r^j$ and $\tau^{(N)}$: 
\bea(ll)
 V (x_q, y_q)^N  &=   \zeta_0^L X^\frac{N(N+1)}{2} \alpha (\sigma_q)^{-N} S(\sigma_q) \prod_{j=1}^N \frac{r_j (t_q, \sigma_q)}{ \tau^{(N)} (\omega^j t_q)}  \\
&= \zeta_0^L (\prod_{k=1}^{N-1} z(\omega^k t)^{-k}) \alpha(\sigma^\dagger )^N S(\sigma_q)   \prod_{j=1}^N  \frac{ \tau^{(N)} (t_q)}{  r_j (t_q, \sigma^\dagger) }  .
\elea(VSrjt)

\subsection{Eigenvalues of a finite-size  transfer matrix $V (q)$ of the chiral Potts model   \label{ssec.EigV}}
In this subsection, we computer the eigenvalue $V (x_q, y_q)$ using the Bethe solutions, $F(t_q)$ and $F'(t_q)$, of equation (\req(Bethet2)) (\req(Bethet2')), where $h^\pm, h'^\mp$ are defined in (\req(hpm)), (\req(h'pm)) respectively.
By (\req(hh'AF)), the $\tau^{(2)}T$-relation (\req(tauV)) can be written as    
\bea(l)
\tau^{(2)}(t_q) V (Uq) =  h^+(t_q)    \frac{(t_{p'}^N -  t_q^N )^{\frac{L}{N}}}{ (t_{p'}-t_q)^L  } V ( q) +  h^-(\omega t_q)  \frac{ (t_{p'} -  \omega t_q)^L }{(t_{p'}^N -  t_q^N )^{\frac{L}{N}}  }  X V (U^2 q), 
\\
\tau^{(2)}(t_q) V (U'q) =    h'^-(t_q)  \frac{(t_{p'}^N - t_q^N )^{\frac{L}{N}}  }{ (t_{p'} -  t_q)^L } X V (q)  +  h'^+(\omega t_q) \frac{ (t_{p'} -  \omega  t_q)^L }{(t_{p'}^N -  t_q^N )^{\frac{L}{N}}  } V (U'^2 q). 
\elea(tVh)
Here the $N$th root-of-unity $\gamma$, $ \gamma'$  in $h^+(t_q)$, $h'^-(t_q)$ are absorbed  in the $\frac{L}{N}$th power of $t_{p'}^N -  t_q^N$. By (\req(VVT)) and (\req(tVh)),  the $\tau^{(2)}T$-relation can also be expressed by
\bea(l)
\tau^{(2)}(t_q) V (U'q) =    h^+(t_q)  \frac{(t_{p'}^N - t_q^N )^{\frac{L}{N}}  }{  (t_{p'} -  t_q)^L }  \omega^P X V (q)  +  h^-(\omega t_q) \frac{  (t_{p'} -  \omega  t_q)^L }{ (t_{p'}^N -  t_q^N )^{\frac{L}{N}}  } \omega^{-P} V (U'^2 q),
\\
\tau^{(2)}(t_q) V (Uq) =  h'^-(t_q)    \frac{(t_{p'}^N -  t_q^N )^{\frac{L}{N}}}{ (t_{p'}-t_q)^L  } \omega^{-P} V ( q) 
+  h'^+(\omega t_q)  \frac{(t_{p'} -  \omega t_q)^L }{ (t_{p'}^N -  t_q^N )^{\frac{L}{N}}  }  \omega^P X V (U^2 q), 
\elea(tVhP)
where  $\omega^P$ is  some $N$th root of unity. Note that for generic $p$ and $p'$, the variable $q$  are restricted is $(t_q, \sigma_q) \in W^+$ for the first relations of (\req(tVh))and (\req(tVhP)), with but  $(t_q, \sigma_q) \in W^-$ for the second ones there. By (\req(tjFt)), $r_j(t, \sigma)$ in (\req(rj)) and $\tau^{(N)}(t)$  are related by (\cite{MR} (30abc)):
\bea(l)
r_j(t, \sigma) = 
 \omega^{-j(Q+P_a)}h^+(t)\cdots h^+(\omega^{j-1}t) \frac{F( t )}{ F(\omega^j t )} \tau^{(N)}(\omega^j t) , \\
r_j(t, \sigma^\dagger) = \omega^{jP_a} h^-(t)h^-(\omega t) \cdots h^- (\omega^{j-1} t) \frac{F( \omega^j t )}{F(t )} \tau^{(N)}(t)
\elea(rjtN)
for $j=1, \ldots, N$,
where $(t, \sigma) \in W^+$. The product of formulas  in (\req(rjtN)), together with (\req(hpd)), yields
$$
\begin{array}{l}
\prod_{j=1}^N r_j(t, \sigma) = 
 \omega^\frac{-N(N+1)(Q+P_a)}{2}\frac{\alpha_q^N F( t )^N}{(\prod_{k=1}^{N-1} h^+(\omega^k t)^k )(\prod_{j=1}^N F(\omega^j t ))} \prod_{j=1}^N\tau^{(N)}(\omega^j t), \\
\prod_{j=1}^N r_j(t, \sigma^\dagger) =   \omega^\frac{N(N+1)P_a}{2} \frac{\overline{\alpha}_q^N \prod_{j=1}^N F( \omega^j t )}{(\prod_{k=1}^{N-1} h^-(\omega^k t)^k ) F(t )^N } \tau^{(N)}(t)^N .
\end{array}
$$
Substituting either one of the above relations into (\req(VSrjt)), one finds the expression of $V (x_q, y_q)$ :
$$
V (x_q, y_q) =   \zeta_0^\frac{L}{N}    
 \frac{F( t_q )}{\prod_{k=1}^{N-1} (t_{p'} - \omega^k t_q)^\frac{kL}{N}}     
\bigg( \frac{S(\sigma_q) \prod_{k=1}^{N-1} (t_{p'} - \omega^k t_q)^{kL} }{(\prod_{k=1}^{N-1} h^+(\omega^k t_q)^k) (  \prod_{j=1}^N F(\omega^j t_q )) }\bigg)^\frac{1}{N} 
\omega^\frac{-(N+1)P_a}{2} .
$$
By the properties of $S(\sigma)$ in (\req(Vavg)) and $F(t)$ in (\req(Fpol)), the analysis  of power orders near points where $t=0$ or $t=\infty$ with a finite $y$-value in turn yields the expression  
\be
\bigg(\frac{S(\sigma_q) \prod_{k=1}^{N-1} (t_{p'} - \omega^k t_q)^{kL} }{(\prod_{k=1}^{N-1} h^+(\omega^k t_q)^k) (  \prod_{j=1}^N F(\omega^j t_q )) }\bigg)^\frac{1}{N} 
\omega^\frac{-(N+1)P_a}{2} = x_q^{P_x} y_q^{P_y} \mu_q^{-P_\mu} G( \sigma_q )
\ele(G)
where $P_x, P_y$ and $P_\mu$ are integers with $P_\mu \equiv 0 \pmod{N}$, and $G( \sigma_q )$ is an algebraic function of $\sigma_q$ with $(t_q, \sigma_q) \in W^+$ and a finite value except where $F(t_q)= 0$ or $\sigma = \sigma_p^\dagger$. 
To avoid the ambiguity of assigning integers $P_x, P_y$ in case $W_1^{\prime \prime \prime}$ where $x_q$ differs from $y_q$ by a constant factor, for the rest of this subsection, we shall assume the rapidity curve not equal to $W_1^{\prime \prime \prime}$, i.e.  $W= W_{k'}, W_1^\prime, W_1^{\prime \prime}$. The integers $P_x, P_y, P_\mu$ are chosen so that 
$G( 0 ) \neq 0$ with $G( \sigma_q )$ taking non-zero value when $t_q=0$, ($P_\mu$ is defined to be zero in the case $W= W_1^{\prime \prime}$). Hence
\be
V (x_q, y_q) =   \zeta_0^\frac{L}{N} x_q^{P_x } y_q^{P_y}  \mu_q^{-P_\mu}   
 \frac{F( t_q )}{\prod_{k=1}^{N-1} (t_{p'} - \omega^k t_q)^\frac{kL}{N}}  G( \sigma_q ), ~ ~ 
(t_q, \sigma_q) \in W^+.
\ele(VFG) 
By the consistency of  Bethe relation (\req(t2F)) and $\tau^{(2)}V$-relations (\req(tVh)) (\req(tVhP)), the integers $P_x, P_y$ satisfy $P_x \equiv P_a  , P_y \equiv P_a +Q + P  \pmod{N}$ where $P$ is in (\req(tVhP)). 
Similarly, by (\req(tjFt')) one finds  
\bea(l)
r_j(t, \sigma) = \omega^{-j P_b} h'^-( t)\cdots h'^-(\omega^{j-1} t)  \frac{F'( t )}{F'(\omega^j t )} \tau^{(N)}(\omega^j t) , \\
r_j(t, \sigma^\dagger) = \omega^{j(P_b-Q)} (\prod_{k=0}^{j-1} h'^+(\omega^k t)) \frac{F'(\omega^j t )}{ F'( t )} \tau^{(N)}(t) , 
\elea(rj'tN)
for $(t, \sigma) \in W^-$, which in turn yields 
\be
V (x_q, y_q) =   \zeta_0^\frac{L}{N} x_q^{P'_x} y_q^{P'_y } \mu_q^{-P'_\mu}   
 \frac{F'( t_q )}{\prod_{k=1}^{N-1} (t_{p'} - \omega^k t_q)^\frac{kL}{N}}  G'( \sigma_q ), \ \ 
(t_q, \sigma_q) \in W^-,
\ele(VFG')
where $P'_x \equiv P_b  -Q - P , P'_y \equiv P_b   \pmod{N}$, and $G'(\sigma)$ is defined by
\be
\bigg(\frac{S(\sigma_q) \prod_{k=1}^{N-1} (t_{p'} - \omega^k t_q)^{kL} }{(\prod_{k=1}^{N-1} h'^-(\omega^k t_q)^k) (  \prod_{j=1}^N F'(\omega^j t_q )) }\bigg)^\frac{1}{N} 
\omega^\frac{-(N+1)(P_b-Q)}{2} = x_q^{P'_x} y_q^{P'_y } \mu_q^{-P'_\mu} G'( \sigma_q )
\ele(G')
such that $G'( 0^\dagger ) \neq 0$ and $G'( \sigma_q )$ is non-zero when $t_q=0$. Therefore one obtains
$$
P_x + P'_y \equiv P_y + P'_x \equiv P_a+P_b  .
$$
Note that (\req(VFG)) and (\req(VFG')) are valid on $W^+$ and $W^-$ respectively, but the formulas are not true on the whole Riemann surface $W$  in general when the vertical rapidities  $p, p'$ are arbitrary. One needs both relations, (\req(VFG)) and (\req(VFG')), for an equivalent expression of the $N$th $TT$-relation in (\req(VTT)):
\bea(ll)
D y^{Nm} \mu_q^{-Nd}
G( \sigma_q )  G'( \sigma^\dagger_q ) &=  t_q^{-(P_x + P'_y) } \frac{\tau^{(N)} (t_q)}{F( t_q )F'( t_q )} \prod_{k=1}^{N-1} (\frac{y_p^2 y_{p'}^2( t_{p'} - \omega^k t_q) }{\omega \mu_p \mu_{p'}(t_p - \omega^k t_q)})^\frac{kL}{N} , \\ 
D x^{-Nm} \mu_q^{-Nd}
G( \sigma_q )  G'( \sigma^\dagger_q )  & =  t_q^{-( P'_x+ P_y ) } \frac{\tau^{(N)} (t_q)}{F( t_q )F'( t_q )} \prod_{k=1}^{N-1} (\frac{y_p^2 y_{p'}^2( t_{p'} - \omega^k t_q) }{\omega \mu_p \mu_{p'}(t_p - \omega^k t_q)})^\frac{kL}{N} , 
\elea(TTFF')
where $(t_q, \sigma_q) \in W^+ $,  and $d, m$ are integers with $N d := P_\mu -P'_\mu $, and 
$N m= P'_x+ P_y - (P_x + P'_y ) ~ \equiv 0 \pmod{N}$. One can also use the variable $(t_q, \sigma_q) \in W^-$  to  the express (\req(TTFF'))  by interchanging $\sigma_q, x_q, \mu_q$ with $\sigma^\dagger_q, y_q, \mu_q^{-1}$ respectively.  The right hand sides of (\req(TTFF')) are algebraic functions of $t$, invariant when replacing $t$ by $\omega t$, and they define the same function when $m = 0$. It is known that $y_q^N$ is a never-vanishing function on $W^+$ when $W= W_{k'} ~ (|k'|< 1)$,  $W_1^\prime, W_1^{\prime \prime}$; while $x_q^N$ is non-zero on $W^+$ when  $W= W_{k'} ~ (|k'|> 1)$. Correspondingly, we denote the $t$-function of the right hand side in (\req(TTFF')) by $P(t)$. Then $P(0) \neq 0$, and $P(t)$ is indeed a $t^N$-function. In the case $W_1^\prime$, one finds $m=0$ in (\req(TTFF')), where two relations define the same function $P(t)$. By adjusting the integers $P_a$ and $P_b$, one may assume $P_x = P_a $, $P'_y = P_b $ in the first relation of (\req(TTFF')), or $P_y =  P_a  +Q + P$, $P'_x = P_b  -Q - P$ in second one. Hence $P(t)$ is expressed by 
\be
P (t) = t_q^{-P_a-P_b  } \frac{\tau^{(N)} (t)}{F( t )F'( t )} \prod_{k=1}^{N-1} (\frac{y_p^2 y_{p'}^2( t_{p'} - \omega^k t) }{\omega \mu_p \mu_{p'}(t_p - \omega^k t)})^\frac{kL}{N} 
\ele(P)
with $P(0) \neq 0$. The above $P (t)$ is related to the functions, $G$ and $G'$, by 
\bea(ll)
D y^{Nm} \mu_q^{-Nd}
G( \sigma_q )  G'( \sigma^\dagger_q ) =  P(t_q), & (t_q, \sigma_q) \in W^+_{k'} ~ (|k'|< 1),  W_1^{\prime +}, W_1^{\prime \prime +}; \\ 
D x^{-Nm} \mu_q^{-Nd}
G( \sigma_q )  G'( \sigma^\dagger_q )   =  P(t_q), & (t_q, \sigma_q) \in W^+_{k'} ~ (|k'|> 1), W_1^{\prime +}. 
\elea(GGtN)
Note that $d=0$ in the case $W_1^{\prime \prime}$ where $P_\mu, P'_\mu $ are defined to be zero. In the case $W= W_{k'}$ or $W_1^\prime$, $\sigma_q = \mu^N_q$, and the $\infty$-limit of $t_q$ in $W^+$ corresponds to $\infty$-limit of $x_q^N$ with a finite $y_q^N$-value. When $t_q \rightarrow \infty$, one finds $d \geq 0$ in the first relation of  (\req(GGtN)), and $d-m \geq 0$ in the second one. Note that as functions of $W$, the $t^N$-function $P(t)$ in (\req(P)) can be decomposed as a product function : $P(t) = {\cal G}(\sigma){\cal G}(\sigma^\dagger) $, where ${\cal G}(\sigma)$ is a $\sigma$-function for $\sigma \in \CZ$. However, the functions $G(\sigma), \mu_q^{-Nd} G'(\sigma^\dagger)$ in (\req(TTFF')) can not be expressed as ${\cal G}(\sigma)$, ${\cal G}(\sigma^\dagger)$ in general (for arbitrary $p', p'$). Indeed, say when $m=0$, the relation (\req(GGtN)) implies that there exists some factorization of $P(t) = {\cal G}(\sigma){\cal G}(\sigma^\dagger) $ such that $G (\sigma ) = {\cal G} (\sigma ) g(t, \sigma), \mu^{Nd} G' (\sigma ) = {\cal G} (\sigma ) g'(t, \sigma)$, where $g(t, \sigma)$ and $g'(t, \sigma)$ are never-vanishing functions on $ W^+, W^-$ respectively such that $D g(t, \sigma)g'(t, \sigma^\dagger) = 1$ for $(t, \sigma) \in W^+$. 

For the rest of this subsection, we determine the quantum numbers $P_a, P_b$ in the case of generic $p$ and $p'$. We leave the special, i.e. the  alternating superintegrable (\req(alsup)), case in the next subsection. 
By the $L$-operator expression (\req(L)) with parameters in (\req(tpp')), the eigenvalue $\tau^{(2)}(t)$ is a $t$ -polynomial of degree $L$ with the leading and constant terms given by
\be
\lim_{t \rightarrow \infty} \frac{\tau^{(2)}(t)}{t^L} = (\frac{-\omega }{y_p y_{p'}})^L ( 1 + \omega^Q \mu_p^L \mu_{p'}^L), \ ~ \ ~ \tau^{(2)}(0) = 1 + \omega^Q z(0). 
\ele(t2lc)
On the other hand, $\tau^{(2)}(t)$  is expressed by 
(\req(t2F)) and(\req(t2F')), where $F(t), F'(t)$ are the Bethe solution of (\req(Bethet2)), (\req(Bethet2')) respectively.  
Using (\req(hpml)), (\req(h'pml)), (\req(hpd)) and (\req(h'pd)), one obtains another expression of the leading and constant terms of $\tau^{(2)}(t)$:
\bea(ll)
\lim_{t \rightarrow \infty} \frac{ \tau^{(2)} (t)}{t^L}&= (\frac{-\omega}{y_p y_{p'}})^L (\omega^{-P_a+ P_\gamma -J-L} + \omega^{Q+ P_a-P_\gamma +L + J  } \mu_p^L \mu_{p'}^L ) \\
&= (\frac{-\omega}{y_p y_{p'}})^L(  \omega^{Q-P_b+ P_{\gamma'} -J' -L } ( \mu_p^N \mu_{p'}^N)^\frac{L}{N} + \omega^{P_b-P_{\gamma'} +L + J'}   \mu_p^L \mu_{p'}^L ( \mu_p^N \mu_{p'}^N)^\frac{-L}{N}  ) 
, \\
\tau^{(2)}(0) &= \omega^{-P_a}    + \omega^{Q+P_a} z(0) =  \omega^{P_b} + \omega^{Q-P_b}  z(0)   .
\elea(t2l0) 
For generic $p, p'$ in (\req(tpp')) when both $z(0)$ and $\mu_p^L \mu_{p'}^L$ not being $N$th roots of unity, the relations, (\req(t2lc)) and  (\req(t2l0)), imply the following constraints of the quantum numbers in (\req(Bethet2)) and (\req(Bethet2')):
\be
P_a \equiv P_b \equiv 0, \ ~ ~ J  \equiv P_\gamma -L ,  \ ~ ~  J' \equiv  P_{\gamma'}-L + r .  
\ele(Pab0)
where $r$ is the integer determined by $\mu_p^L \mu_{p'}^L  = ( \mu_p^N \mu_{p'}^N)^\frac{L}{N} \omega^{-r}$.
In this situation, $h^\pm (t)$ in (\req(hpm)) is defined only on $W^+$,  while $h'^\mp (t)$ in (\req(h'pm)) defined only on $W^-$. By the discussion in section \ref{ssec.Bethe},  $F(t)$ in  (\req(Bethet2)) and $F'(t)$ in  (\req(Bethet2')) are different polynomials. The eigenvalues of chiral Potts transfer matrix defined by (\req(G)) and (\req(VFG)) are in one-to-one correspondence with eigenvalues (\req(t2F)) of $\tau^{(2)}$-matrix obtained by the Bethe equation (\req(Bethet2)). Hence the $\tau^{(2)}$-matrix is non-singular when the rapidity parameters  $p, p'$ in (\req(tpp')) of the $L$-operator (\req(L)) are generic. The corresponding chiral Potts eigenvalues are expressed by (\req(VFG)) and (\req(VFG')), which satisfy the relation (\req(tVh)).

\subsection{Eigenvalues $V(q)$ for the alternating superintegrable chiral Potts model \label{ssec.AlSup}}
The alternating superintegrable case\footnote{Here we use a slightly general notion of alternating superintegrability than that in \cite{B93} (6.1), \cite{B94} (14) or \cite{BBP} section  6, where the alternating superintegrable CPM is when $x_{p'}=y_p$, $y_{p'}= x_p$. Our general setting includes one special case of XXZ chains associated to cyclic $U_q(sl_2)$ representations with both $q$ and the representation parameter $\varsigma$ being $N$th roots of unity \cite{R075}.}  
 is when the vertical rapidities $p, p'$ in (\req(tpp')) satisfy one of the following equivalent relations:
\be
x_p^N = y_{p'}^N \Longleftrightarrow y_p^N = x_{p'}^N  \Longleftrightarrow \sigma_p = \sigma_{p'}^\dagger  \Longleftrightarrow \alpha_q^\frac{1}{L} = \overline{\alpha}_q^\frac{1}{L} 
\ele(alsup)
where the variables $\sigma, \sigma^\dagger$ are in (\req(sigma)), and $\alpha_q, \overline{\alpha}_q ~ (q \in W)$ are functions in (\req(aaq)). When the rapidity curve is ${\goth W}_{k'}$ in (\req(xymu)) or ${\goth W}_1^\prime $ in (\req(Cdeg)), the above conditions are equivalent to $\lambda_p \lambda_{p'}=1$. In the alternating superintegrable case, the function $\frac{\tau^{(N)}(t)}{F(t)^2}$ with the Bethe solution $F(t)$ of (\req(Bethet2)) is a $t$-polynomial: $\frac{\tau^{(N)}(t)}{F(t)^2} \in \CZ [t]$. 
Indeed by (\req(tjFt)), $\frac{\tau^{(N)}(t)}{F(t)^2}$ is regular except the zeros $t_0$'s of $F(t)$. By (\req(hpd)), the finite-valued condition of $\frac{\tau^{(N)}(t)}{F(t)^2}$  at $t_0$'s is equivalent to the vanishing of $ \frac{ \overline{\alpha}_q }{\alpha_q }
 h^+(\omega^{-1} t) F(\omega^{-1} t)  + \omega^{(Q+2P_a) } h^-(t)F(\omega t) $ at $t_0$'s, which is the same as the Bethe condition  (\req(Bethet2)) when $ \alpha_q = \overline{\alpha}_q $. In this section, we discuss the alternating superintegrable case. Write 
\be
x_p = \omega^{\rm m} y_{p'}, ~ ~ x_{p'}= \omega^{\rm m'} y_p , ~ ~  \mu_p\mu_{p'} = \omega^{\rm n} ~ ~ ~ ~ ({\rm m}, {\rm m'}, {\rm n} \in \ZZ) ,
\ele(xyAS)
and define the variable ${\tt t}:=  \frac{t}{ y_p y_{p'}}$. Since $e^\pm (t) =1$, the functions
$h^\pm (t), h'^\mp (t) $ in (\req(hpm)), (\req(h'pm)) are 
\bea(ll)
h^+(t)= {\tt h}^+({\tt t}) = (1 -\omega^{-{\rm m'}} {\tt t})^L , & h^-(t)= {\tt h}^-({\tt t}) = \omega^{(1+ {\rm m} + {\rm m'}+{\rm n})L} (1 - \omega^{-{\rm m}}{\tt t})^L ; \\
h'^-(t) = {\tt h}'^- ({\tt t}) = \omega^{ (1+{\rm m + m'+ n })L } (1 -\omega^{-{\rm m'}} {\tt t})^L,  &h'^+(t) = {\tt h}'^+ ({\tt t})= (1 - \omega^{-{\rm m}}{\tt t})^L  
\elea(Suphh)
with $P_\gamma = -{\rm m'}L $ and $P_{\gamma'} = (1+{\rm m + n })L $. The above ${\tt t}$-polynomials are considered as functions on $W$. Using the relation between ${\tt h}^\pm$ and ${\tt h}'^\mp$,  the expressions (\req(t2F)) and (\req(t2F')) for $\tau^{(2)}(t)$  are the same with $F(t)= F'(t)$ and $
P_b - P_a \equiv Q +(1+{\rm m}+{\rm m'} +{\rm n})L \pmod{N}$. The $\omega^P$ in (\req(tVhP)) is given by $P = P_{\gamma'}- P_{\gamma}$. Write $F(t)$ in (\req(Fpol)) as $F(t) = {\tt F}({\tt t}) = \prod_{j=1}^J (1+ \omega {\tt v}_j {\tt t})$ where $ {\tt v}_j= y_p y_{p'} v_j $. 
Both (\req(Bethet2)) and (\req(Bethet2')) define the same Bethe equation:
\bea(l)
( \frac{{\tt v}_i +\omega^{ -{\rm m}-1}  }{{\tt v}_i + \omega^{-{\rm m'}-2} } )^L  = - \omega^{-P_a-P_b} \prod_{j=1}^J \frac{{\tt v}_i -  \omega^{-1}   {\tt v}_j }{ {\tt v}_i -\omega  {\tt v}_j } , \ \ i= 1, \ldots, J. 
\elea(ASBe)
Hence $\tau^{j}(t)$  in (\req(tjFt)) and (\req(tjFt')) is expressed by
$$
\begin{array}{l}
\tau^{(j)}(t)
= \omega^{(j-1)P_b } {\tt F}({\tt t}) {\tt F}(\omega^j {\tt t} ) 
\sum_{n=0}^{j-1} \frac{\{(1 -\omega^{-{\rm m'}} {\tt t}) \cdots (1 -\omega^{-{\rm m'}+n-1} {\tt t}) (1 - \omega^{-{\rm m}+n+1}{\tt t}) \cdots (1 - \omega^{-{\rm m}+j-1}{\tt t})\}^L \omega^{-n(P_a+P_b) } }{ 
 {\tt F}(\omega^n {\tt t})  {\tt F}(\omega^{n+1}{\tt t})}.
\end{array}
$$
In particular, one obtains the $t$-polynomial 
\bea(l)
\frac{\tau^{(N)}(t)}{{\tt F}(t)^2 }
=\omega^{-P_b}  
\sum_{n=0}^{N-1} \frac{\{ (1 -\omega^{-{\rm m'}} {\tt t}) \cdots (1 -\omega^{-{\rm m'}+n-1} {\tt t})  (1 - \omega^{-{\rm m}+n+1}{\tt t}) \cdots (1 - \omega^{-{\rm m}+N-1}{\tt t}) \}^L \omega^{-n(P_a+P_b)} }{ 
 {\tt F}(\omega^n {\tt t})  {\tt F}(\omega^{n+1}{\tt t})}. 
\elea(AStN)
Now (\req(G)) and (\req(G')) are defined on $W$ with $P_x=P_x', P_y = P_y'$. The functions  $G, G'$ (by a proper choice of phase-factors) are related by
$$
G(\sigma) =  \mu^{Nd} G'(\sigma) , ~ ~ \ N d := P_\mu -P'_\mu .
$$
The eigenvalue in (\req(VFG)), equivalently (\req(VFG')), becomes 
\be
V (x_q, y_q) =   \zeta_0^\frac{L}{N} x_q^{P_a } y_q^{P_b }  \mu_q^{-P_\mu}   
 \frac{F( t_q )}{\prod_{k=1}^{N-1} (t_{p'} - \omega^k t_q)^\frac{kL}{N}}  G( \sigma_q ), \ \ 
(t_q, \sigma_q) \in W.
\ele(VFGAS) 
where $G(\sigma)$ in (\req(GGtN)) and the $t^N$-function  $P(t)$ in (\req(P)) are expressed in the following form: 
\bea(c)
 D  G( \sigma )  G( \sigma^\dagger ) =  P(t), ~ ~ ~  (t, \sigma) \in W , \\
P (t) := C {\tt t}^{ -P_a -P_b }
 \prod_{k=1}^{N-1} (\frac{( 1 - \omega^{-{\rm m'} +k} {\tt t}) }{(1 - \omega^{-{\rm m}+ k} {\tt t})})^\frac{kL}{N} \frac{\tau^{(N)} (t)}{{\tt F}( {\tt t} )^2 }, \ \ P(0) \neq 0  , 
\elea(GGtNAS)
with $C= \omega^{\frac{({\rm m'}-{\rm m} -{\rm n} -1) (N-1)L}{2}} (y_p y_{p'})^{-P_a-P_b +(N-1)L}$. The non-vanishing of $G(\sigma)$ and $G( \sigma^\dagger )$ when $t_q=0$ implies the integers $P_a, P_b$ satisfying 
\be
0 \leq P_a + P_b \leq N-1 , \ \ P_b - P_a \equiv Q +(1+{\rm m}+{\rm m'} +{\rm n})L \pmod{N}. 
\ele(PabAS)
Note that the relations, (\req(VFGAS)) and (\req(GGtNAS)), are still valid  when the rapidity curve is $W_1^{\prime \prime \prime}$ in the alternating superintegrable case. By (\req(t2lc)), (\req(t2l0) ) and  $z(0)= \omega^{(1+{\rm n}+{\rm m} + {\rm m'})L}$, one finds 
$$
1 + \omega^{P_b-P_a} =  \omega^{P_b} + \omega^{-P_a}, ~ ~  1 + \omega^{Q+{\rm n}L}  = \omega^{-P_a-{\rm m'}L  -J-L} + \omega^{Q+ P_a+ {\rm m'}L +L + J  +{\rm n}L},
$$
which imply
\be
P_a \equiv 0 ~ {\rm or} ~ P_b \equiv 0 ,  ~ ~ \ \ J +P_b 
\equiv ({\rm m}+ {\rm n}) L  +Q , ~  ({\rm m}+2 {\rm m'} )L  \pmod{N}. 
\ele(PabAsp)

The parameters in the $L$-operator (\req(L)) of  $\tau^{(2)}$-models equivalent to the alternating superintegrable case satisfy the relations: ${\sf c}^N=1$, ${\sf a}^N= {\sf b'}^N, {\sf a'}^N= {\sf b}^N$ (\cite{R0710} (2.23)). These models are characterized by $\omega^{\rm m}, \omega^{\rm m'}, \omega^{\rm n}$ in (\req(xyAS))  with the $L$-operator
\be
{\tt L} ( {\tt t} ) = \left( \begin{array}{cc}
        1  -  \omega^{\rm n} {\tt t}  X   & (1  -\omega^{{\rm m} +{\rm n}+1} X) Z \\
       - {\tt t} ( 1  -   \omega^{{\rm m'} + {\rm n}}  X )Z^{-1} & - {\tt t} + \omega^{{\rm m} + {\rm m'} + {\rm n}+1}  X
\end{array} \right) .
\ele(Lalsup)
The chiral Potts transfer matrix $T_{p, p'}(q)$ with $p, p'$ in (\req(xymu)) or (\req(Cdeg)) can be  constructed from the $\tau^{(2)}$-matrix as a Baxter's $Q$-operator (see \cite{BBP, BazS} or \cite{R0710} section 3.1). For each $\tau^{(2)}$-eigenvalue, one associates the $t^N$-function  $P(t)$ in (\req(GGtN)). Each factor function $G(\sigma)$ of $P(t)$ gives rise to an eigenvalue (\req(VFG)) of $T_{p, p'}(q)$. Hence the degeneracy of $\tau^{(2)}$-model occurs, with a power of 2 degenerate states for each $\tau^{(2)}$-eigenvalue. Note that the $Q$-operators of the same $\tau^{(2)}$-matrix in this context are distinct for different choices of rapidity curves. It is known \cite{R075} when $N$ is odd: $N= 2M+1$, the $\tau^{(2)}$-model  with $\omega^{{\rm m}+{\rm m'}+1}= \omega^{\rm n}=1$ is  identified with XXZ chains associated to cyclic $U_q(sl_2)$ representations with both $q$ and the representation parameter $\varsigma$ being $N$th roots of unity. In these cases, the $\tau^{(2)}$-degeneracy carries  the $sl_2$-loop-algebra symmetry. In particular, when ${\rm m}= {\rm m}' =M$, the $\tau^{(2)}$-model is the spin-$\frac{N-1}{2}$ XXZ chain, with the $Q$-operator identified with  the homogeneous  chiral Potts transfer matrix for $p= p'$:  $(x_p, y_p, \mu_p) = (\omega^M \eta^\frac{1}{2}, \eta^\frac{1}{2} , 1)$, where $\eta$ is in (\req(eta)). In this case, the $sl_2$-loop-algebra symmetry of XXZ chain inherits the Onsager-algebra-symmetry induced from the Hamiltonian chain of the chiral Potts model \cite{R05o}. 

The homogeneous superintegrable chiral Potts model is the case (\req(alsup)) when $p=p' \not\in W^{\prime \prime \prime}$   with the coordinates
$$
p= p'  ~ ~ : (x_p, y_p, \mu_p) = (\omega^{{\rm m}+ k } \eta^\frac{\pm 1}{2}, \omega^{k} \eta^\frac{\pm 1}{2} , \omega^\frac{\rm n}{2}) , 
$$
where $\eta$ is in (\req(eta)). The $\tau^{(N)}(t), P(t)$ in (\req(AStN)) and (\req(GGtN)) respectively are polynomials  expressed by 
\bea(l)
\tau^{(N)}(t)= C^{-1} {\tt t}^{P_a+P_b} {\tt F}({\tt t})^2 P(t), \
P (t ) =  C \omega^{-P_b}  
\sum_{n=0}^{N-1} \frac{(1 -{\tt t}^N)^L  (\omega^n{\tt t})^{-(P_a+P_b)} }{ (1- \omega^{n-{\rm m}} {\tt t})^L {\tt F}(\omega^n {\tt t})  {\tt F}(\omega^{n+1}{\tt t})} ,
\elea(homsP)
where $C= \omega^{\frac{-( {\rm n} + 1) (N-1)L}{2}} (\omega^{2k} \eta^{\pm 1})^{-P_a-P_b +(N-1)L}$, and  $P_a, P_b$ are integers in (\req(PabAS)) with ${\rm m'}= {\rm m}$.
Indeed, $P (t)$ is a ${\tt t}^N$-polynomial which defines the evaluation polynomial of the Onsager-algebra symmetry in the homogeneous superintegrable chiral Potts model \cite{R05o}. The homogeneous superintegrable CPM with rapidities in ${\goth W}_{k'}$ and ${\rm m}= k = {\rm n}= 0$ has been discussed extensively in literature, and the eigenvalues of chiral Potts transfer matrix are given in \cite{AMP} \cite{B89, B91, B90, B93, B94}: 
$$
T(q) = N^L \frac{(\eta^{\frac{-1}{2}} x_q-1)^L}{(\eta^{-N/2} x_q^N-1)^L} (\eta^{\frac{-1}{2}}x_q)^{P_a}(\eta^{\frac{-1}{2}}y_q)^{P_b}\mu_q^{-P_\mu} \frac{F(\eta^{-1} t_q) }{\omega^{P_b}F(\omega \eta)} {\cal G} (\lambda_q) 
$$
where ${\cal G}(\lambda_q)$ satisfies ${\cal G} (\lambda_q) {\cal G} (\lambda_q^{-1}) = \frac{P(t)}{P(\eta)}$. By (\req(Vdef)), one finds
$$
V(x_q, y_q)= \frac{ e^{\frac{\pi {\rm i}(N-1)(N-2)L}{12N}}N^\frac{L}{2} \eta^\frac{-P_a-P_b}{2}}{\omega^{P_b} F(\omega \eta )} x_q^{P_a}y_q^{P_b}\mu_q^{-P_\mu}  \frac{ F(\eta^{-1} t_q) }{ \prod_{k=1}^{N-1} (1 - \omega^k \eta^{-1} t_q)^\frac{kL}{N}} {\cal G} (\lambda_q) .
$$  
Then the eigenvalue $S(\lambda_q)$ in (\req(Vavg)) is expressed by 
$$
(\frac{S(\lambda_q)}{\prod_{k=1}^NF(\omega^k \eta^{-1} t_q)})^\frac{1}{N} =  \frac{ N^\frac{L}{2} \eta^\frac{-P_a-P_b}{2} }{\omega^{P_b+ \frac{(N-1)L}{4}}F(\omega \eta)}  x_q^{P_a}y_q^{P_b}\mu_q^{-P_\mu} {\cal G} (\lambda_q) \omega^\frac{P_a (N-1)}{2} . 
$$
Hence $G(\lambda_q)$ relates to ${\cal G} (\lambda_q)$ by 
$$
G( \lambda_q ) = e^{\frac{-\pi {\rm i}(N-1)L}{2N}} N^\frac{L}{2} \frac{ \eta^{\frac{1}{2}(-P_a-P_b+ L(N-1))} }{\omega^{P_b}F(\omega \eta)}  {\cal G} (\lambda_q)
$$
so that  (\req(G)), (\req(VFG)) and (\req(VFGAS))  hold. The evaluation polynomial $P(t)$ is defined by (\req(homsP)) with  ${\rm m}= k = {\rm n}= 0$. The relation (\req(GGtN)) holds for $D= \frac{\omega^{P_b}F(\omega \eta )}{ F(\eta)  } (= e^{{\rm i} P})$, where $P$ is the total momentum (\cite{BBP} (4.49), \cite{AMP} (2.24))\footnote{Here we use the convention in \cite{BBP}, where the total momentum  differs from that in \cite{AMP} (2.24) or \cite{R05o} (62)  by a minus sign.}. The quantum numbers, $P_a, P_b$ and $J$, are the integers in (\req(PabAS)) (\req(PabAsp)), now expressed by
\bea(ll)
(i) & 0 \leq P_a + P_b \leq N-1 , \ \ P_b - P_a \equiv Q + L \pmod{N} , \\
(ii) & P_a \equiv 0 ~ {\rm or} ~ P_b \equiv 0 , \ \ J +P_b \equiv 0, ~ Q   \pmod{N}.
\elea(Pabhsp)
The above $(i)$ is the condition \cite{B93} (6.16),  \cite{B94} (24) for $r=0$, and  \cite{AMP} (C.4) in the case  $N=3$. The condition $(ii)$ could be known for the specialists as it is supported by the numerical studies for $N=3$ in Table I of \cite{AMP}, but it seems not appeared in literature before to the best of the author's knowledge.  This novel constraint is consistent with the total momentum $e^{{\rm i} P L}= 1$ condition: $LP_b \equiv J( Q-2P_b -J) \pmod{N}$ (\cite{AMP} (C.3), \cite{R05o} (63)), and it also agrees with the conclusion, $P_a  \equiv 0$ or $ P_b \equiv 0$, in the algebraic Bethe ansatz discussion of $\tau^{(2)}$-model (\cite{R06F} section 3.2). 

\section{Concluding Remarks}\label{sec.F} 
In this work, we study the eigenvalue spectrum of a finite-size transfer matrix of the chiral Potts model with alternating rapidities by use of the functional relations in \cite{BBP}. Here the rapidity curve is either ${\goth W}_{k'}$ in (\req(xymu)) \cite{BPA}, or  a curve in (\req(Cdeg)) for the degenerate models, which include the selfdual solutions of the star-triangle relation (\req(TArel)) \cite{AMPT, AuP, FatZ, MPTS}. We first establish the Bethe equations, (\req(Bethet2)) and (\req(Bethet2')), of $\tau^{(2)}$-model through the Wiener-Hopf splitting (\req(hpm)) (\req(h'pm)) of $\alpha_q$ and $\overline{\alpha}_q$ as in \cite{MR}.
We express the $\tau^{(j)}$-eigenvalues by using the Bethe solution, then obtain the expression, (\req(VFG)), (\req(VFG')) together with (\req(GGtN)), of eigenvalues  $V(x_q, y_q)$  of the finite transfer matrix in CPM. The procedure enables one to 
solve the eigenvalue problem of CPM of a finite size. In the alternating superintegrable case (\req(alsup)), the $\tau^{(2)}$-models for all $k'$ are the same, produced by the $L$-operator (\req(Lalsup)) with a simple form of Bethe equation (\req(ASBe)). As in the homogeneous superintegrable CPM \cite{R05o}, the degeneracy of the $\tau^{(2)}$-matrix occurs, and the chiral Potts transfer matrix serves a $Q$-operator of the $\tau^{(2)}$-matrix. The eigenvalues, (\req(VFGAS)) and (\req(GGtN)), of the chiral Potts transfer matrix share a similar structure as those of homogeneous superintegrable CPM in \cite{AMP, B93, B94}. We find the derivation in this paper has enhanced our structural understanding about the natures of eigenvalues in CPM. The explicit calculation will lead to some physical implications as compared with previous works in homogeneous chiral Potts model in \cite{AMP, B89, MR}. In this work, we study the CPM mainly from the aspect of the hermitian quantum chain case as in \cite{MR}. The physics of the quantum spin chain Hamiltonian is different from the statistical model with the positive Boltzmann weight case, which was the main  concern in \cite{AJP, B91,B03}, (see, e.g. \cite{AuP}). These two cases overlap only in the critical limiting case of \cite{FatZ}. In this paper the issues about the  peculiarities of those various physical regimes were not discussed. Here, just to keep things simple, we restrict our attention only to the mathematical aspect of eigenvalue spectrum of the finite-size transfer matrix. We leave the physical discussion of ours results, and possible generalizations to future work.

\section*{Acknowledgements} 
The author is pleased to acknowledge the hospitality of Laboratoire de Mathematiques et Physique Theorique, CNRS/UMR, University of Tours, France (2007, fall), and U C Berkeley, U.S.A. (2008, spring),  where parts of this work were carried out. He wishes to thank Professor P. Baseilhac, and Professor S. Kobayashi for their invitation.
This work is supported in part by National Science Council of Taiwan under Grant No NSC 96-2115-M-001-004.

\end{document}